\documentclass[aps,prc,twocolumn,groupedaddress,showpacs,superscriptaddress,nofootinbib]{revtex4-1}
\usepackage{graphicx,amsmath}
\usepackage{amssymb}
\usepackage{hyperref}
\usepackage[version=3]{mhchem}
\begin{document}


\title{Thermonuclear fusion rates for tritium + deuterium using Bayesian methods}



\author{Rafael S. de Souza}\email{drsouza@ad.unc.edu}
\affiliation{Department of Physics \& Astronomy, University of North Carolina at Chapel Hill, NC 27599-3255, USA}
\author{S. Reece Boston}
\affiliation{Department of Physics \& Astronomy, University of North Carolina at Chapel Hill, NC 27599-3255, USA}
\author{Alain Coc}
\affiliation{Centre de Sciences Nucl\'eaires et de Sciences de la Mati\`ere, Univ. Paris-Sud, CNRS/IN2P3, Universit\'e Paris-Saclay, B\^atiment, 104, F-91405 Orsay Campus, France}
\author{Christian Iliadis}\email{iliadis@unc.edu}
\affiliation{Department of Physics \& Astronomy, University of North Carolina at Chapel Hill, NC 27599-3255, USA}
\affiliation{Triangle Universities Nuclear Laboratory (TUNL), Durham, North Carolina 27708, USA}

\date{\today}

\begin{abstract}
The $^3$H(d,n)$^4$He reaction has a large low-energy cross section and will likely be utilized in future commercial fusion reactors. This reaction also takes place during big bang nucleosynthesis. Studies of both scenarios require accurate and precise fusion rates. To this end, we implement a one-level, two-channel R-matrix approximation into a Bayesian model. Our main goals are to predict reliable astrophysical S-factors and to estimate R-matrix parameters using the Bayesian approach. All relevant parameters are sampled in our study, including the channel radii, boundary condition parameters, and data set normalization factors. In addition, we take uncertainties in both measured bombarding energies and S-factors rigorously into account. Thermonuclear rates and reactivities of the $^3$H(d,n)$^4$He reaction are derived by numerically integrating the Bayesian S-factor samples. The present reaction rate uncertainties at temperatures between $1.0$~MK and $1.0$~GK are in the range of 0.2\% to 0.6\%. Our reaction rates differ from previous results by 2.9\% near 1.0~GK. Our reactivities are smaller than previous results, with a maximum deviation of 2.9\% near a thermal energy of $4$~keV. The present rate or reactivity uncertainties are more reliable compared to previous studies that did not include the channel radii, boundary condition parameters, and data set normalization factors in the fitting. Finally, we investigate previous claims of electron screening effects in the published $^3$H(d,n)$^4$He data. No such effects are evident and only an upper limit for the electron screening potential can be obtained. 
\end{abstract}


\maketitle


\section{Introduction} \label{sec:intro}
The cross section of the $^3$H(d,n)$^4$He reaction has a large Q-value of $17.6$~MeV, and a large cross section that peaks at $\approx$ $5$~barn near a deuteron (triton) bombarding energy of $105$~keV ($164$~keV). For these reasons, the $^3$H(d,n)$^4$He reaction will most likely fuel the first magnetic and inertial confinement fusion reactors for commercial energy production \cite{wilson08,kim12}. The reactors are expected to operate in the thermal energy range of kT$=$ $1$ $-$ $30$~keV, corresponding to temperatures of T$=$ $12$ $-$ $350$~MK. These values translate to kinetic energies between $4$~keV and $120$~keV in the $^3$H $+$ $d$ center-of-mass system, which can be compared with a Coulomb barrier height of $\approx$ $280$~keV. Accurate knowledge of the $^3$H(d,n)$^4$He thermonuclear rate is of crucial importance for the design of fusion reactors, plasma diagnostics, fusion ignition determination, and break-even analysis. The $^3$H(d,n)$^4$He reaction also occurs during big bang nucleosynthesis, at temperatures between $0.5$~GK and $1.0$~GK, corresponding to center-of-mass Gamow peak energies in the range of $13$ $-$ $252$~keV.

The $^3$H $+$ $d$ low-energy cross section is dominated by a s-wave resonance with a spin-parity of J$^\pi$ $=$ $3/2^+$, corresponding to the second excited level near E$_x$ $\approx$ $16.7$~MeV excitation energy in the $^5$He compound nucleus \cite{tilley02}. This level decays via emission of d-wave neutrons. It has mainly a $^3$H $+$ $d$ structure, corresponding to a large deuteron spectroscopic factor \cite{barker97}, while shell model calculations predict a relatively small neutron spectrocopic factor \cite{barker85}. However, the neutron penetrability is much larger than the deuteron penetrability at these low energies, so that incidentally the partial widths for the deuteron and neutron channel ($\Gamma_d$, $\Gamma_n$), given by the product of spectroscopic factor and penetrability, become similar in magnitude. This near equality of the deuteron and neutron partial widths causes the large low-energy cross section of the $^3$H(d,n)$^4$He reaction \cite{conner52,argo52} since, considering a simple Breit-Wigner expression, the cross section maximum is proportional to $\Gamma_d \Gamma_n / (\Gamma_d + \Gamma_n)^2$, which peaks for the condition $\Gamma_d$ $\approx$ $\Gamma_n$. 

Different strategies to analyze the data have been adopted previously. Fits of the available $^3$H(d,n)$^4$He data using Breit-Wigner expressions were reported by Duane \cite{duane72} and Angulo {\it et al.} \cite{angulo99}, while a Pad\'e expansion was used by Peres \cite{peres79}. Single-level and multi-level R-matrix fits to $^3$H(d,n)$^4$He data were discussed by Jarmie, Brown and Hardekopf \cite{jarmie84}, Brown, Jarmie and Hale \cite{brown87}, Barker \cite{barker97}, and Descouvemont {\it et al.} \cite{descouvemont04}. A comprehensive R-matrix approach that included elastic and inelastic cross sections of the $^3$H $+$ $d$ and $^4$He $+$ $n$ systems in addition to the $^3$H(d,n)$^4$He data, incorporating $2664$ data points and $117$ free parameters, was presented by Hale, Brown and Jarmie \cite{hale87} and Bosch and Hale \cite{bosch92,bosch92b}. An analysis of  $^3$H(d,n)$^4$He data using effective field theory, with only three fitting parameters, can be found in Brown and Hale \cite{brown14}.

Our first goal is to quantify the uncertainties in the thermonuclear rates and reactivities for the $^3$H(d,n)$^4$He reaction. All previous works employed chi-square fitting in the data analysis, assuming Gaussian likelihoods throughout, and disregarding any uncertainties in the center-of-mass energies. Here, we will discuss an analysis using Bayesian techniques. This approach has major advantages, as discussed by Iliadis {\it et al.} \cite{iliadis16} and G\'omez I\~nesta, Iliadis and Coc \cite{gomez17}, because it is not confined to the use of Gaussian likelihoods, and instead allows for implementing those likelihoods into the model that best apply to the problem at hand. Also, all previous R-matrix analyses kept the channel radii and boundary condition parameters constant during the fitting. In reality, these quantities are not rigidly constrained, and their variation will impact the uncertainties of the derived S-factors and fusion rates. Furthermore, uncertainties affect not only the measured S-factors, but also the experimental center-of-mass energies. Uncertainties in both independent and dependent variables can be easily implemented into a Bayesian model, whereas no simple prescription for such a procedure exists in chi-square fitting. Our second goal is to investigate the usefulness of the Bayesian approach for estimating R-matrix parameters. The results will prove useful in future studies that involve multiple channels and resonances.

In Section~\ref{sec:data}, we briefly present the S-factor data adopted in the present analysis. Section~\ref{sec:rmatrix} summarizes the reaction formalism. Bayesian hierarchical models are discussed in Section~\ref{sec:bayes}, including likelihoods, model parameters, and priors. Section~\ref{sec:fixparam} considers some preliminary ideas. Our Bayesian model for the $^3$H(d,n)$^4$He reaction is presented in Section~\ref{sec:tdnmodel}. Results are presented in Section~\ref{sec:results}. In Section~\ref{sec:rates}, we present Bayesian reaction rates and reactivities. A summary and conclusions are given in Section~\ref{sec:summary}. An evaluation of the data adopted in our analysis is presented in Appendix~\ref{app:data}.

\section{Data Selection} \label{sec:data}
Several previous works have used all available $^3$H $+$ $d$ cross section data in the fitting. A rigorous data analysis requires a careful distinction between statistical and systematic uncertainties (Section~\ref{sec:uncert}), because we aim to implement these effects separately in our Bayesian model. For this reason, we will consider only those experiments for which we can quantify the two contributions independently. Detailed information regarding the experimental uncertainties is provided in Appendix~\ref{app:data}. 

The $^3$H(d,n)$^4$He low-energy cross section represents a steep function of energy. For example, at $20$~keV in the center of mass, an energy variation of only $0.1$~keV causes a 2\% change in  cross section, while at $10$~keV a variation of $0.1$~keV causes a 6\% change in the cross section. Therefore, accurate knowledge of the incident beam energy becomes crucial for predicting cross sections and thermonuclear rates. Experiments that employed thin targets will be less prone to systemic effects than those using thick targets. For example, consider the data measured by Argo {\it et al.} \cite{argo52}, which were adopted at face value in previous fusion rate determinations. Argo {\it et al.} \cite{argo52} employed $1.5$~mg/cm$^2$ thick aluminum entrance foils for their deuterium gas target. Under such conditions, tritons that slowed down to a laboratory energy of $183$~keV after passing the entrance foil would have lost $568$~keV in the foil, giving rise to an overall beam straggling of about $31$~keV. In this case, it is difficult to reliably correct the cross section for the beam energy loss. Compare this situation to the measurement by Jarmie and collaborators \cite{jarmie84,jarmie84b}, where the triton beam lost an energy less than $200$~eV while traversing a windowless deuterium gas target. A detailed discussion of all data sets that have been adopted or disregarded in the present analysis is given in Appendix~\ref{app:data}.
 
\begin{figure}[]
\includegraphics[width=1.0\columnwidth]{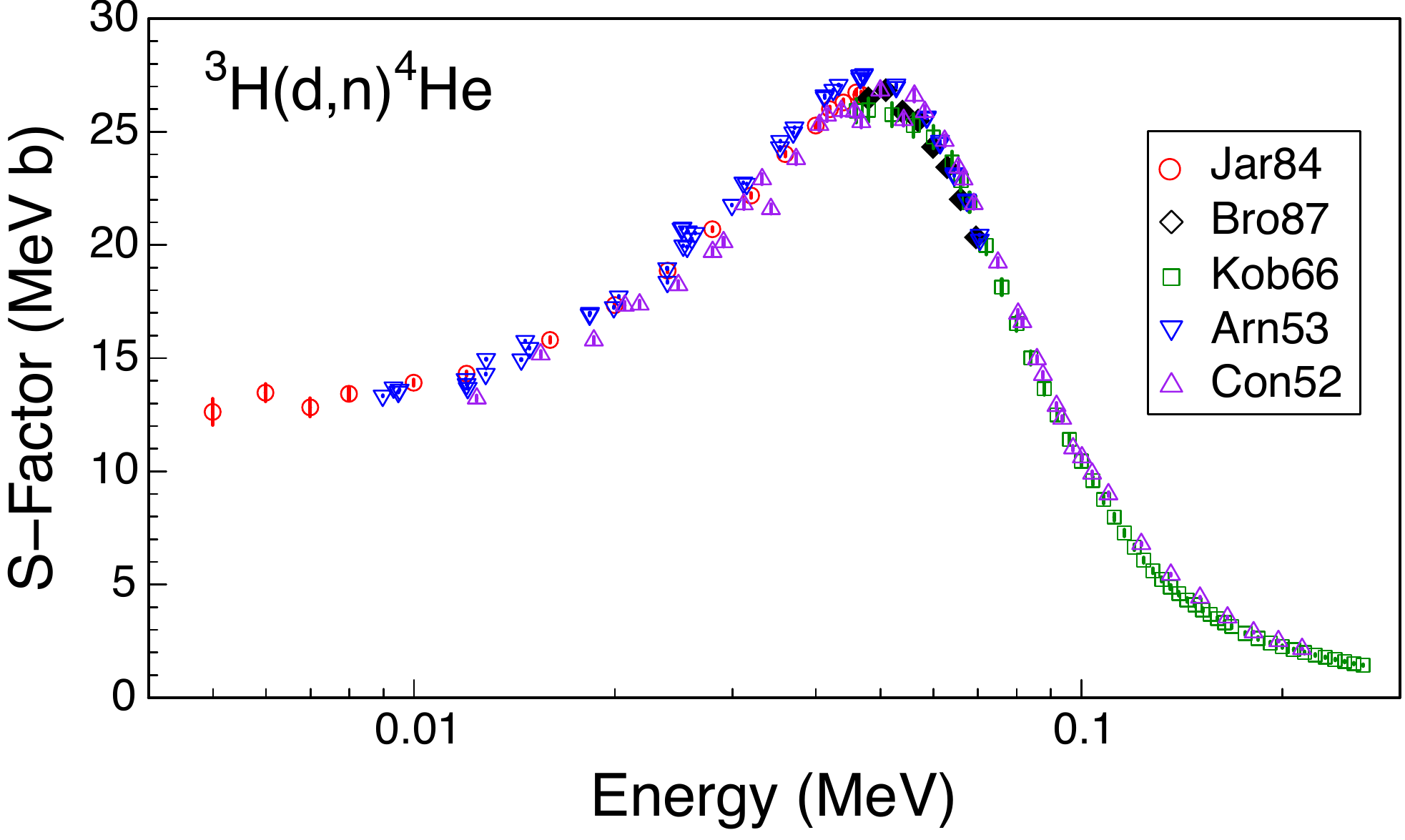}
\caption{\label{fig:data} The data used in our analysis: (Red circles) Jarmie, Brown and Hardekopf \cite{jarmie84};  (Black diamonds) Brown, Jarmie and Hale \cite{brown87}; (Green squares) Kobzev, Salatskij and Telezhnikov \cite{kobzev66}; (Blue triangles) Arnold {\it et al.} \cite{arnold53}; (Purple triangles) Conner, Bonner and Smith \cite{conner52}. Absolute cross sections were not determined in Ref.~\cite{brown87} and their data were normalized to those of Ref.~\cite{jarmie84}. Only statistical uncertainties are shown, but for many of the data points they are smaller than the symbol sizes. Details regarding the data evaluation are given in Appendix~\ref{app:data}. The energy ranges important for fusion reactors and big bang nucleosynthesis are $4$ $-$ $120$~keV and $13$ $-$ $252$~keV, respectively.
}
\end{figure}

All of our adopted data are shown in Figure~\ref{fig:data}. They originated from the experiments by Jarmie, Brown and Hardekopf \cite{jarmie84}, Brown, Jarmie and Hale \cite{brown87}, Kobzev, Salatskij and Telezhnikov \cite{kobzev66},  Arnold {\it et al.} \cite{arnold53}, and Conner, Bonner and Smith \cite{conner52}, and contain $191$ data points in the center-of-mass energy region between $5$~keV and $270$~keV. Notice that the results of Ref.~\cite{brown87} have been used at face value in previous fusion rate estimations, although these authors did not determine any absolute cross sections. In Section~\ref{sec:tdnmodel} we will discuss how to implement such data  into a Bayesian model.

\section{Reaction Formalism}\label{sec:rmatrix}
Since we are mainly interested in the low-energy region, where the $3/2^+$ s-wave resonance dominates the cross section, we will describe the theoretical energy dependence of the cross section using a one-level, two-channel R-matrix approximation. This assumption is justified by previous works that found that the measured S-factor data are about equally well reproduced by single-level and multi-level  R-matrix analyses at center-of-mass energies below $\approx$ $100$~keV (see, e.g., Figure 4 in Ref.~\cite{hale14}).

The angle-integrated cross section of the $^3$He(d,n)$^4$He reaction is given by
\begin{equation}
\sigma_{dn}(E) = \frac{\pi}{k^2} \frac{2J+1}{(2j_1 + 1)(2j_2 + 1)} \left| S_{dn} \right|^2 \label{eq:eq1}
\end{equation}
where $k$ and $E$ are the wave number and energy, respectively, in the $^3$H $+$ $d$ center-of-mass system, $J$ $=$ $3/2$ is the resonance spin, $j_1$ $=$ $1/2$ and $j_2$ $=$ $1$ are the spins of the triton and deuteron, respectively, and $S_{dn}$ is the scattering matrix element. The corresponding astrophysical S-factor is defined by
\begin{equation}
S_{bare}(E)  \equiv E e^{2\pi\eta} \sigma_{dn}(E) \label{eq:eq2}
\end{equation}
where $\eta$ is the Sommerfeld parameter. The scattering matrix element for a single level can be expressed as \cite{lane58}
\begin{equation}
\left| S_{dn} \right|^2 = \frac{\Gamma_d \Gamma_n}{(E_0 + \Delta - E)^2 + (\Gamma/2)^2} \label{eq:scatt}
\end{equation}
where $E_0$ denotes the level eigenenergy. The partial widths of the $^3$H $+$ $d$ and $^4$He $+$ $n$ channels ($\Gamma_d$, $\Gamma_n$), the total width ($\Gamma$), and total level shift ($\Delta$), which are all energy dependent, are given by 
\begin{equation}\label{eq:gamwidth}
\Gamma = \sum_c \Gamma_c = \Gamma_d + \Gamma_n~,~~~\Gamma_c = 2 \gamma_c^2 P_c
\end{equation}
\begin{equation}
\Delta = \sum_c \Delta_c = \Delta_d + \Delta_n~,~~~\Delta_c = - \gamma_c^2 (S_c - B_c) \label{eq:shift}
\end{equation}
where $\gamma_c^2$ is the reduced width\footnote{ In this work, we are not using the Thomas approximation \cite{thomas51}. Therefore, our partial and reduced widths are ``formal'' R-matrix parameters. Use of the Thomas approximation necessitates the definition of ``observed'' R-matrix parameters, which has led to significant confusion in the literature.}, and $B_c$ is the boundary condition parameter. The energy-dependent quantities $P_c$ and $S_c$ denote the penetration factor and shift factor, respectively, for channel $c$ (either $^3$H $+$ $d$ or $^4$He  $+$ $n$). They are computed numerically from the Coulomb wave functions, $F_\ell$ and $G_\ell$, according to 
\begin{equation}
P_c = \frac{k a_c}{F_\ell^2 + G_\ell^2}~,~~~S_c = \frac{k a_c(F_\ell F_\ell^\prime+G_\ell G_\ell^\prime)}{F_\ell^2 + G_\ell^2}
\end{equation}
The Coulomb wave functions and their radial derivatives are evaluated at the channel radius, $a_c$, and the quantity $\ell$ denotes the orbital angular momentum for a given channel.

In some cases, the fit to the data can be improved by adding a distant level in the analysis, located at a fixed energy outside the range of interest. However, such ``background poles'' have no physical meaning. As will become apparent below, the single-level, two-channel approximation represents a satisfactory model for the low-energy data of interest here.

Teichmann and Wigner \cite{teichmann52} showed that the reduced width, $\gamma^2_{\lambda c}$, of an eigenstate $\lambda$ cannot exceed, {\it on average}, the single-particle limit, given by
\begin{equation}
\left< \gamma^2_{\lambda c} \right> \lesssim \frac{3}{2}\frac{\hbar^2}{m_c a_c^2} \label{eq:teich}
\end{equation}
where $m_c$ is the reduced mass of the interacting particle pair in channel $c$. In this original formulation, Equation~(\ref{eq:teich}) only holds for a reduced width that is averaged over many eigenstates, $\lambda$. Using the actual strength of the residual interaction in nuclei, Dover, Mahaux and Weidenm\"uller \cite{dover69} found a single-particle limit of
\begin{equation}
\gamma^2_{\lambda c} \lesssim \frac{\hbar^2}{m_c a_c^2} \label{eq:wl}
\end{equation}
for an individual resonance in a {\it nucleon} channel. The quantity $\gamma^2_{WL}$ $\equiv$ $\hbar^2/(m_c a_c^2)$ is often referred to as the Wigner limit. Considering the various assumptions made in deriving the above expressions, the Wigner limit provides only an approximation for the maximum value of a reduced width. The Wigner limit can also be used to define a ``dimensionless reduced width'', $\theta_{\lambda c}^2$, according to
\begin{equation}
\gamma^2_{\lambda c} \equiv \frac{\hbar^2}{m_c a_c^2} \theta_{\lambda c}^2 \label{eq:wl2}
\end{equation}

We perform the S-factor fit to the data using the expression \cite{assenbaum87,engstler88}
\begin{equation}
S(E) \approx S_{bare}(E) e^{\pi\eta (U_e/E)}
\label{eq:SE}
\end{equation}
where $U_e$ is the energy-independent electron screening potential. The latter quantity has a positive value and depends on the identities of target and projectile, i.e., it differs for forward and inverse kinematics experiments.

R-matrix parameters and cross sections derived from data have a well-known dependence on the channel (or interaction) radius, which is usually expressed as
\begin{equation}
a_c = r_0 \left(  A^{1/3}_{1} +  A^{1/3}_{2}  \right)
\end{equation}
where $A_i$ are the mass numbers of the interacting nuclei, and $r_0$ is the radius parameter, with a value usually chosen between $1.4$~fm and $1.5$~fm. The channel radius dependence arises from the truncation of the R-matrix to a restricted number of poles (i.e., a finite set of eigenenergies). The radius of a given channel has no rigorous physical meaning, except that the chosen value should exceed the sum of the radii of the colliding nuclei (see, e.g., Descouvemont and Baye \cite{descouvemont10}, and references therein). The radius dependence can likely be reduced by including more levels (including background poles) in the data analysis, but only at the cost of increasing the number of fitting parameters. In any case, it is important to include the effects of varying the channel radius in the data analysis. We will address this issue in Section~\ref{sec:tdnmodel}.

Another point that needs investigating is the effect of the arbitrary choice of the boundary condition parameter, $B_c$. It can be seen from Equations~(\ref{eq:scatt}) and (\ref{eq:shift}) that changing $B_c$ will result in a corresponding change of the eigenenergy, $E_0$, to reproduce the measured location of the cross section maximum. Lane and Thomas \cite{lane58} recommended to chose $B_c$  in the one-level approximation such that the eigenvalue $E_0$ lies within the width of the measured resonance. 

For a relatively narrow resonance, one can assume that the measured location of the cross section (or S-factor) maximum, $E_r$, coincides with the maximum of the scattering matrix element, which occurs when the first term in the denominator of Equation~(\ref{eq:scatt}) is set equal to zero. In that case, the resonance energy, $E_r$, can be defined by 
\begin{equation}\label{eq:narrow}
E_0 + \Delta(E_r) - E_r = 0 
\end{equation}
One (but not the only) choice for the boundary condition parameter is then $B_c$ $=$ $S_c(E_r)$. This choice results in $\Delta(E_r)$ $=$ $0$, or $E_r$ $=$ $E_0$, in agreement with the recommendation of Lane and Thomas \cite{lane58}. This procedure, which represents the standard assumption in the literature, cannot be easily applied in the case of the exceptionally broad low-energy resonance in $^3$H(d,n)$^4$He, as will be discussed in Section~\ref{sec:fixparam}.

\section{Bayesian Inference}\label{sec:bayes}
\subsection{General Aspects}\label{sec:genaspec}
We analyze the S-factor data using Bayesian statistics and Markov chain Monte Carlo (MCMC) algorithms. The application of this method to nuclear astrophysics is discussed in Iliadis {\it et al.} \cite{iliadis16} and G\'omez I\~nesta, Iliadis and Coc \cite{gomez17}. Bayes' theorem is given by \cite{jaynes03}
\begin{equation}
p(\theta|y) = \frac{\mathcal{L}(y|\theta)\pi(\theta)}{\int \mathcal{L}(y|\theta)\pi(\theta)d\theta}
\label{eq:bayes}
\end{equation}
where the data are denoted by $y$ and the complete set of model parameters is described by the vector $\theta$. All factors entering in Equation~(\ref{eq:bayes}) represent probability densities: $\mathcal{L}(y|\theta)$ is the likelihood, i.e., the probability that the data, $y$, were obtained assuming given values of the model parameters; $\pi(\theta)$ is called the prior, which represents our state of knowledge about each parameter before seeing the data; the product of likelihood and prior defines the posterior, $p(\theta|y)$, i.e., the probability of the values of a specific set of model parameters given the data; the denominator, called the evidence, is a normalization factor and is not important in the context of the present work. It can be seen from Equation~(\ref{eq:bayes}) that the posterior represents an update of our prior state of knowledge about the model parameters once new data become available.

The random sampling of the posterior is usually performed numerically over many parameter dimensions using MCMC algorithms \cite{metropolis53,hastings70,geyer11}. A Markov chain is a random walk, where a transition from state $i$ to state $j$ is independent (�memory-less�) of how state $i$ was populated. The fundamental theorem of Markov chains states that for a very long random walk the proportion of time (i.e., the probability) the chain spends in some state $j$ is independent of the initial state it started from. This set of limiting, long random walk, probabilities is called the stationary (or equilibrium) distribution of the Markov chain. When a Markov chain is constructed with a stationary distribution equal to the posterior, $p(\theta | y)$, the samples drawn at every step during a sufficiently long random walk will closely approximate the posterior density. Several related algorithms (e.g., Metropolis, Metropolis-Hastings, Gibbs) are known to solve this problem numerically. The combination of Bayes� theorem and MCMC algorithms allows for computing models that are too difficult to estimate using chi-square fitting.

In this work, we use a MCMC sampler based on the differential evolution adaptive Metropolis (DREAM) algorithm \cite{terbraak08,laloy12}. This method employs multiple Markov chains in parallel and uses a discrete proposal distribution to evolve the sampler to the posterior density. It has been shown to perform well in solving complex high-dimensional search problems. This sampler is implemented in the ``BayesianTools'' package, which can be installed within the R language \cite{rcore15}. Running a Bayesian model refers to generating random samples from the posterior distribution of model parameters. This involves the definition of the model, likelihood, and priors, as well as the initialization, adaptation, and monitoring of the Markov chains.

\subsection{Types of uncertainties}\label{sec:uncert}
Of particular interest for the present work is the concept of a hierarchical Bayesian model (see Hilbe, de Souza and Ishida \cite{hilbe17}, and references therein). It allows us to take all relevant effects and processes into account that affect the measured data, which is often not possible with chi-square fitting. We first need to define the different types of uncertainties impacting both the measured energy and S-factor in a nuclear physics experiment.

{\it Statistical} (or {\it random}) uncertainties usually follow a known probability distribution. When a series of independent experiments is performed, statistical uncertainties will give rise to different results in each individual measurement. Statistical uncertainties can frequently be reduced by improving the data collection procedure or by collecting more data. They have a number of different causes. For example, for the S-factor, one source is the Poisson uncertainty, which derives from measuring $N$ counts with an associated uncertainty of $\sqrt{N}$. Another source is caused by the background that needs to be subtracted from the measured total intensity to find the net intensity of the signal. A third source is introduced by the detector, which is subject to additional random uncertainties (e.g., corrections for detection efficiencies). The cumulative effect causes the measured number of signal counts to fluctuate randomly from data point to data point. 

{\it Systematic} uncertainties originate from sources that systematically shift the signal of interest either too high or too low. They do not usually signal their existence by a larger fluctuation of the data, and they are not reduced by combining the results from different measurements or by collecting more data. When the experiment is repeated, the presence of systematic effects may not produce different answers. Reported systematic uncertainties are at least partially based on assumptions made by the experimenter, are model-dependent, and follow vaguely known probability distributions \cite{heinrich07}. In a nuclear physics experiment, systematic effects impact the overall normalization by shifting all points of a given data set into the same direction. They are correlated from data point to data point, in the sense that if one happened to know how to correct such an uncertainty for one data point, then one could calculate the correction for the other data points as well.

In many cases, the scatter about the best-fit model is larger than can be explained by the reported measurement uncertainties. It its useful in such situations to introduce an {\it extrinsic} uncertainty, which describes additional sources of uncertainty in the data that were not properly accounted for by the experimenter. For example, the reported statistical uncertainties may have been too optimistic because target thickness or ion beam straggling effects were underestimated; or perhaps systematic effects that impact data points {\it differently} in a given experiment were unknown to the experimenter.  

To summarize, we assume that three independent effects impact the measured energies and S-factors: (i) statistical uncertainties, which perturb the true (but unknown) energy or S-factor by an amount of $\epsilon_{stat}$; (ii) systematic uncertainties, which perturb the energy or S-factor by an amount of $\epsilon_{syst}$; and (iii) extrinsic scatter, which perturbs the energy or S-factor by an amount of $\epsilon_{extr}$. The overall goal is to estimate credible values for the true energy and S-factor based on the measured data.

\subsection{Likelihoods and Priors}\label{sec:likeprior} 
For illustrative purposes, we will explain in this section how to construct a hierarchical Bayesian model by focussing on uncertainties in the dependent variable, i.e., the S-factor. Our full Bayesian model, including uncertainties in both energy and S-factor, will be discussed in a later section.

Suppose first that the experimental S-factor, $S^{exp}$, is subject to experimental statistical uncertainties only ($\epsilon_{extr}$ $=$ $\epsilon_{syst}$ $=$ $0$; $\epsilon_{stat}$ $\neq$ $0$). Then the likelihood is given by
\begin{equation}
\mathcal{L}(S^{exp}|\theta) = \prod_{i=1}^N\frac{1}{\sigma_{stat,i}\sqrt{2\pi}}e^{-\frac{\left[ S^{exp}_i - S(\theta)_i\right]^2}{2\sigma_{stat,i}^2}}  \label{eq:like2}
\end{equation}
where $S(\theta)_i$ is the model S-factor (e.g., obtained from R-matrix theory); the product runs over all data points, labeled by $i$. The likelihood represents a product of normal distributions, each with a mean of $S(\theta)_i$ and a standard deviation of $\sigma_{stat,i}$, given by the experimental statistical uncertainty of datum $i$. In symbolic notation, the above expression can be abbreviated by
\begin{equation}
S^{exp}_i \sim N(S(\theta)_i, \sigma_{stat,i}^2) 
\end{equation}
where $N$ denotes a normal probability density, and the symbol ``$\sim$'' stands for ``sampled from.'' If, on the other hand, only extrinsic uncertainties impact the S-factor data ($\epsilon_{syst}$ $=$ $\epsilon_{stat}$ $=$ $0$; $\epsilon_{extr}$ $\neq$ $0$), and we assume that these follow a normal probability distribution with a standard deviation of $\sigma_{extr}$, the likelihood can be written as
\begin{equation}
\mathcal{L}(S^{exp}|\theta) = \prod_{i=1}^N\frac{1}{\sigma_{extr}\sqrt{2\pi}}e^{-\frac{\left[ S^{exp}_i - S(\theta)_i\right]^2}{2\sigma_{extr}^2}}  \label{eq:like1}
\end{equation}
In symbolic notation, we obtain
\begin{equation}
S^{exp}_i \sim N(S(\theta)_i, \sigma_{extr}^2) 
\end{equation}
When both effects are taken simultaneously into account ($\epsilon_{extr}$ $\neq$ $0$; $\epsilon_{stat}$ $\neq$ $0$), the overall likelihood is given by a nested (and cumbersome explicit) expression. In the convenient symbolic notation, we can write
\begin{align}
S^\prime_i & \sim N(S(\theta)_i, \sigma_{stat,i}^2) \\
S^{exp}_i & \sim N(S^\prime_i, \sigma_{extr}^2) 
\end{align}
The last two expressions show in an intuitive manner how the overall likelihood is constructed: first, statistical uncertainties, quantified by the standard deviation $\sigma_{stat,i}$ of a normal probability density, perturb the true (but unknown) value of the S-factor at energy $i$, $S(\theta)_i$, to produce a value of $S_i^\prime$; second, the latter value is perturbed, in turn, by the extrinsic uncertainty, quantified by the standard deviation $\sigma_{extr}$ of a normal probability density, to produce the measured value of $S^{exp}_i$. 

The above discussion demonstrates how any effect impacting the data can be implemented in a straightforward manner into a Bayesian hierarchical model. There is nothing special about adopting normal distributions in the example above, which we only chose to explain a complex problem in simple words. As will be seen below, some of the likelihood functions used in the present work are non-normal.

Each of the model parameters, contained in the vector $\theta$, requires a prior distribution. It contains the information on the probability density of a given parameter prior to analyzing the data under consideration. For example, if our model has only one parameter, $\theta$, and if all we know is that the value of the parameter lies somewhere in a region from zero to $\theta_{max}$, we can write in symbolical notation for the prior
\begin{equation}
\theta \sim U(0, \theta_{max})
\end{equation}
where $U$ denotes a uniform probability density. 
 
Normalization factors related to systematic uncertainties represent a special case. For example, a systematic uncertainty of, say, $\pm5\%$, implies that the systematic factor uncertainty is $1.05$. The true value of the normalization factor, $f$, is unknown at this stage, otherwise there would be no systematic uncertainty. However, we do have one piece of information: the expectation value of the normalization factor is unity. If this would not be the case, we would have corrected the data for the systematic effect.

A useful distribution for normalization factors is the lognormal probability density, which is characterized by two quantities, the location parameter, $\mu$, and the spread parameter, $\sigma$. The median value of the lognormal distribution is given by $x_{med}$ $=$ $e^\mu$, while the factor uncertainty, for a coverage probability of 68\%, is $f.u.$ $=$ $e^\sigma$. We will include in our Bayesian model a systematic effect on the S-factor as an informative, lognormal prior with a median of $x_{med}$ $=$ $1.0$ (or $\mu$ $=$ $\ln x_{med}$ $=$ $0$), and a factor uncertainty given by the systematic uncertainty, i.e., in the above example, $f.u.$ $=$ $1.05$ (or $\sigma$ $=$ $\ln f.u.$ $=$ $\ln (1.05)$). The prior is explicitly given by
\begin{equation}
\pi(f_n) = \frac{1}{ \ln (f.u.)_n \sqrt{2 \pi} f_n} 
e^{- \frac{(\ln f_n )^2}{2 [\ln (f.u.)_n]^2}}
\end{equation}
where the subscript $n$ labels the different data sets. We write in symbolic notation
\begin{equation}
f_n \sim LN(0, [ \ln(f.u.)_n ]^2) 
\end{equation}
where $LN$ denotes a lognormal probability density. For more information on this choice of prior, see Iliadis {\it et al.} \cite{iliadis16}. 

Notice that in chi-square fitting, normalization factors are viewed as a systematic shift in the {\it data} (see, for example, Brown and Hale \cite{brown14}). In the Bayesian model, the reported data are not modified. Instead, during the fitting each data set ``pulls'' on the true S-factor curve with a strength inversely proportional to the systematic uncertainty: a data set with a small systematic uncertainty will pull the true S-factor curve more strongly towards it compared to a data set with a large systematic uncertainty.

In the present work, we employ priors that best reflect the physics involved. Depending on the parameter, we use as priors uniform distributions, broad normal densities truncated at zero, narrow normal densities, and log-normal densities.

\section{Preliminary considerations} \label{sec:fixparam}
Although the $^3$H(d,n)$^4$He cross section is dominated at low energies by only a single resonance, any fitting procedure will face a number of interesting problems. 

First, Argo {\it et al.} \cite{argo52} noted that an equally good fit is obtained for two possible solutions of the partial width ratio ($\Gamma_d$/$\Gamma_n$ $>$ $1$ or $<$ $1$), and that it is not possible to chose between them without additional information about the magnitude of the reduced widths $\gamma_d^2$ and $\gamma_n^2$. They also note, however, that the two solutions do not give widely different parameter values since the $\Gamma_d$/$\Gamma_n$ ratio is of order unity.

\begin{figure}[]
\includegraphics[width=1.0\columnwidth]{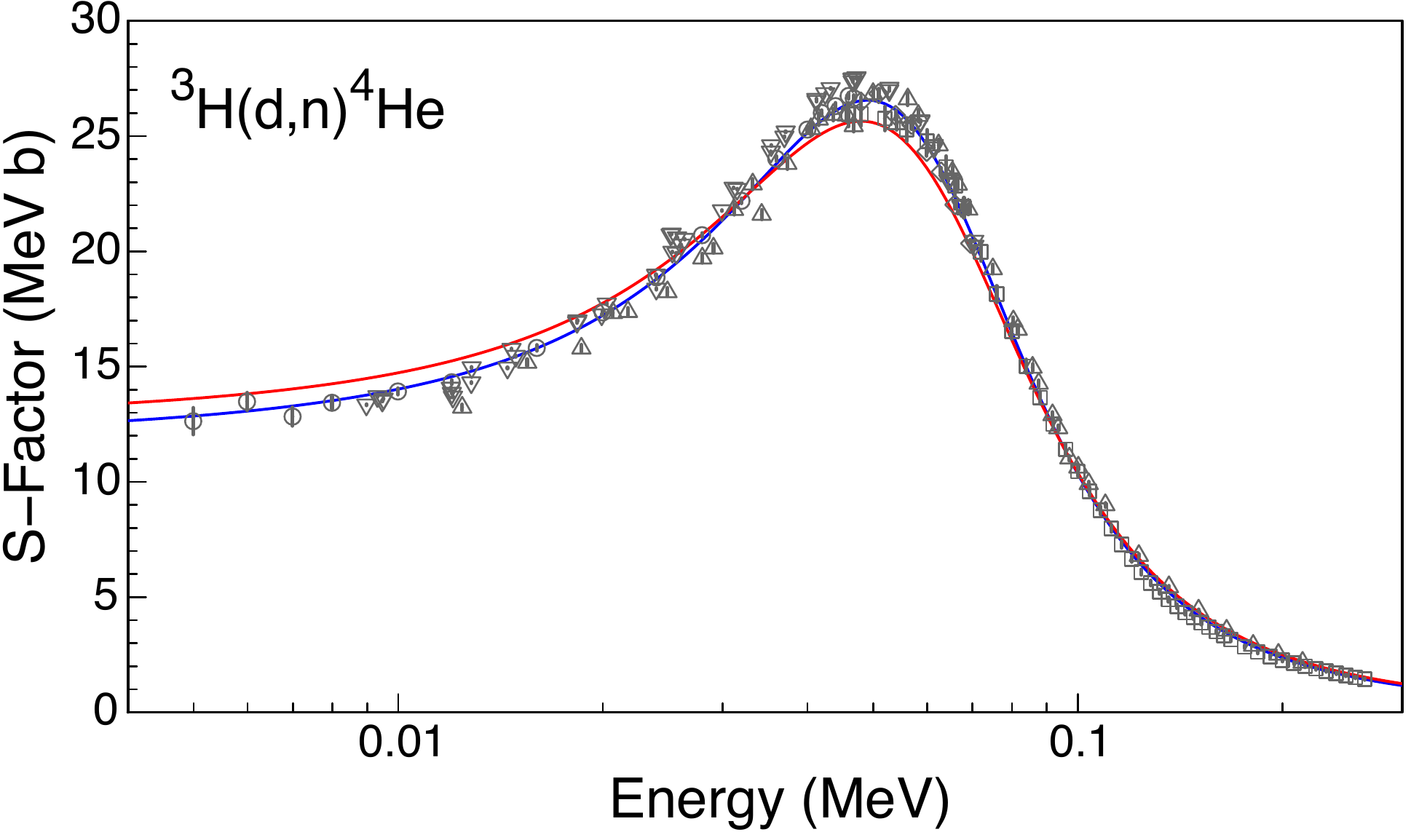}
\caption{\label{fig:prelim} Astrophysical S-factors computed using the single-level, two-channel approximation (see Equations~(\ref{eq:eq1})$-$(\ref{eq:scatt})). The data are the same as in Figure~\ref{fig:data}. The blue curve is computed with the best-fit parameter values of Barker \cite{barker97}. The red curve is obtained by arbitrarily multiplying Barker's reduced widths by a factor of $10$ and adjusting the eigenenergy and boundary condition parameters slightly. The red curve does not represent any best-fit result and serves for illustrative purposes only.
}
\end{figure}

Second, in addition to the ambiguity introduced by the ratio of partial withs, there is another complication related to their absolute magnitude. Consider the two S-factor parameterizations shown in Figure~\ref{fig:prelim}, where the data are the same as in Figure~\ref{fig:data}. The blue curve was obtained using the best-fit values of Barker \cite{barker97} for the eigenenergy and the reduced widths ($E_0$ $=$ $0.0912$~MeV, $\gamma_d^2$ $=$ $2.93$~MeV, $\gamma_n^2$ $=$ $0.0794$~MeV); Barker's fixed values for the channel radii and boundary condition parameters were $a_d$ $=$ $6.0$~fm, $a_n$ $=$ $5.0$~fm, $B_d$ $=$ $-0.285$, $B_n$ $=$ $-0.197$. Barker's derived deuteron reduced width exceeds the Wigner limit by a factor of three, which hints at the exceptional character of the low-energy resonance. Although the data analyzed by Barker and the data evaluated in the present work (see Appendix~\ref{app:data}) are not identical, it can be seen that his best-fit curve (blue) describes the observations well. The red curve was computed by arbitrarily multiplying Barker's reduced width values by a factor of $10$ ($\gamma_d^2$ $=$ $29.3$~MeV, $\gamma_n^2$ $=$ $0.794$~MeV) and slightly adjusting the eigenenergy and boundary condition parameter ($E_0$ $=$ $0.102$~MeV, $B_d$ $=$ $-0.267$). Notice that the red curve does not represent any best-fit result, but its sole purpose is to demonstrate that similar S-factors can be obtained for vastly different values of the partial widths. However, the red curve represents an unphysical result if we consider additional constraints: a deuteron reduced width of $\gamma_d^2$ $=$ $29.3$~MeV, obtained with a channel radius of $a_d$ $=$ $6.0$~fm, exceeds the Wigner limit (see Equation~(\ref{eq:wl})) by a factor of $30$ and is thus highly unlikely.

The latter ambiguity is caused by the structure of Equation~(\ref{eq:scatt}). The large reduced width of the deuteron channel dominates the level shift (see Equation~(\ref{eq:shift})) and also the factor $(E_0 + \Delta - E)$ in Equation~(\ref{eq:scatt}). Therefore, if the reduced or partial widths for both channels are multiplied by a similar factor, the shape and magnitude of the S-factor is only slightly changed. This ambiguity in the parameter selection cannot be removed even when $^3$H $+$ $d$ elastic scattering data are simultaneously analyzed together with the reaction data, as noted by Barit and Sergeev \cite{barit71}.
 
Third, the large total width of the resonance is similar in magnitude to the resonance energy. The resonance is so broad that the experimental values of the scattering matrix element, $|S_{dn}|^2$, the cross section, $\sigma$, and the S-factor, $S_{bare}(E)$, peak at markedly different center-of-mass energies ($\approx$ $80$~keV, $\approx$ $65$~keV, and $\approx$ $50$~keV, respectively). The differences are caused by the energy dependences of the wave number ($k^2$ $\sim$ $E$) in Equation~(\ref{eq:eq1}) and the Gamow factor ($e^{2\pi\eta}$) in Equation~(\ref{eq:eq2}) over the width of the resonance. Furthermore, for given values of $E_0$ and $\Delta$, the location of the $|S_{dn}|^2$ maximum does not coincide anymore with the energy at which the factor $(E_0 + \Delta - E)$ in Equation~(\ref{eq:scatt}) is equal to zero, because of the energy dependence of the penetration factors over the width of the resonance. Therefore, there is no unique procedure for defining an energy, $E_r$, ``at the center of the resonance'' \cite{lane58}, and there is no obvious advantage of adopting the definition of Equation~(\ref{eq:narrow}). In other words, for the exceptionally broad low-energy resonance in $^3$H(d,n)$^4$He, one cannot chose the boundary condition parameter, $B_c$ $=$ $S_c(E_r)$, so that the level shift is zero at the location of  the maximum of either $|S_{dn}|^2$, $\sigma$, or $S_{bare}(E)$, and at the same time expect the ``center of the resonance'', $E_r$, to equal the eigenvalue $E_0$ (see Section~\ref{sec:rmatrix}).\footnote{Jarmie, Brown and Hardekopf \cite{jarmie84} state that they ``chose $B_c$ so that the level shifts $\Delta_c$ are zero near the peak of the S function, which results in the level energy $E_\lambda$ being close to the c.m. energy at which the S function peaks." Their Table VII lists the values of $a_d$ $=$ $5.0$~fm, $a_n$ $=$ $3.0$~fm and $B_d$ $=$ $-0.27864$, $B_n$ $=$ $-0.557$ for the channel radii and boundary conditions, respectively. However, the latter values correspond to an energy of $E_r$ $=$ $90$~keV, which, contrary to their statement, is not near the peak of the astrophysical S factor ($50$~keV).}

For example, consider again the blue curve shown in Figure~\ref{fig:prelim}, which was obtained with $E_0$ $=$ $0.0912$~MeV and $B_d$ $=$ $S_d(E_r)$ $=$ $-0.285$ \cite{barker97}, where the latter value corresponds to an energy of $E_r$ $=$ $0.0912$~MeV. Barker used Equation~(\ref{eq:narrow}) and assumed $E_r$ $=$ $E_0$ in the fitting, but the fitted energies ($E_r$, $E_0$) do not coincide with the measured peak location of the scattering matrix element, or cross section, or S-factor. If we chose instead to set the level shift equal to zero at the location of the $|S_{dn}|^2$ maximum (i.e., $E_r$ $=$ $80$~keV), the eigenenergy needs to be chosen near $152$~keV to achieve a good fit to the data, while keeping all other parameters constant. In other words, the eigenenergy is not located near the $|S_{dn}|^2$ maximum anymore. Conversely, if we set the eigenenergy equal to the location of the maximum of $|S_{dn}|^2$, $\sigma$, or $S_{bare}(E)$, good fits to the data require a level shift of zero near energies of $E_r$ $=$ $0.093$~MeV, $0.097$~MeV, and $0.100$~MeV, respectively. We will explore the impact of boundary condition parameter variations on the fit results in Section~\ref{sec:fixparam}. 

\section{Bayesian model for $^3$H($\lowercase{d}$,$\lowercase{n}$)$^4$H$\lowercase{e}$}\label{sec:tdnmodel}
All previous analyses of the $^3$H(d,n)$^4$He reaction cross section were performed assuming fixed values for the channel radii and boundary condition parameters. However, as explained in Section~\ref{sec:rmatrix}, there is considerable freedom in the choice of these parameters, which, therefore, should be included in the sampling. 

Our model includes the following parameters: (i) R-matrix parameters, i.e., the eigenenergy ($E_0$), reduced deuteron and neutron widths ($\gamma^2_d$, $\gamma^2_n$), deuteron and neutron channel radii ($a_d$, $a_n$), and the boundary condition parameters, $B_c$. (ii) The electron screening potential ($U_e$). (iii) For each of the five data sets, the extrinsic scatter for both energy ($\sigma_{E,extr}$) and S-factor ($\sigma_{S,extr}$), the systematic energy shift ($f_E$), and the S-factor normalization ($f_S$). Overall, our model contains 27 parameters\footnote{Of these $27$ parameters, only $7$ describe uncertainties in the physical model (Equations~\ref{eq:eq2}, \ref{eq:scatt}, and \ref{eq:SE}). The remaining $20$ parameters describe measurement uncertainties, which we introduced for treating the data in our Bayesian model. The large number of the latter parameters does not result in ``overfitting,'' because these parameters are independent of the physical model. In other words, no matter how many measurement uncertainty parameters are introduced in the fitting, our two-channel, single-level R-matrix model will never produce, for example, a double-humped S-factor curve.}.

Normal likelihoods are used for the statistical and extrinsic uncertainties (see also Equations~(\ref{eq:like2}) and (\ref{eq:like1})), because their magnitudes are relatively small. We consider five data sets (Section~\ref{sec:data}), consisting of $191$ data points total. Experimental mean values for the measured energies and S-factors, together with estimates of statistical and systematic uncertainties, are given in Appendix~\ref{app:data}. The priors are discussed next. 

In previous analyses of the $^3$H(d,n)$^4$He reaction cross section, the energy $E_r$ has either been fixed at some arbitrarily value, or the condition $E_r$ $=$ $E_0$ has been arbirarily imposed in the fitting \cite{barker97,coc12}. Neither of these assumptions is justified on fundamental grounds. In Section~\ref{sec:fixparam}, we discussed the complications that arise when choosing the arbitrary value of the boundary condition parameter in the case of a broad resonance. Instead of providing the boundary condition parameters, $B_c$, directly, we find it more useful to report the equivalent results for the energy, $E_B$, at which the level shift is zero according to $B_c$ $=$ $S_c(E_B)$ (see Equation~(\ref{eq:shift})). We use the notation $E_B$ instead of $E_r$ to emphasize that the value of $E_B$ does not correspond to any measured ``resonance energy,'' since such a quantity cannot be determined unambiguously in the present case. Lane and Thomas \cite{lane58} recommended to chose $B_c$ in the one-level approximation such that the eigenvalue $E_0$ lies within the width of the measured resonance. Therefore, we will chose for $E_0$ a uniform prior between $20$~keV and $80$~keV (see Figure~\ref{fig:BaySfac}). For the energy $E_B$, at which the level shift is zero, we adopt a normal density of zero mean value and $1.0$~MeV standard deviation, which is restricted to positive energies only (i.e., a truncated normal density).

Truncated normal densities are also assumed for the reduced widths ($\gamma^2_d$ and $\gamma^2_n$), with standard deviations given by the Wigner limits ($\gamma^2_{WL,d}$ and $\gamma^2_{WL,n}$) for the deuteron and neutron (see Equation~(\ref{eq:wl})). This choice of prior takes into account the approximate character of the Wigner limit concept. For the electron screening potential, we chose a truncated normal density with a standard deviation of $1.0$~keV. 

Descouvemont and Baye \cite{descouvemont10} recommended to chose the channel radius so that its value exceeds the sum of the radii of the colliding nuclei. In a given reaction, the radii of the different channels do usually not have the same value. Previous studies either adopted {\it ad hoc} values, or derived the channel radii from data. Argo {\it et al.} \cite{argo52} and Hale, Brown and Paris \cite{hale14} assumed equal neutron and deuteron channel radii, and find best-fit values of $7.0$~fm from analyzing $^3$H(d,n)$^4$He data. Woods {\it et al.} \cite{woods88} measured the $^4$He($^7$Li,$^6$Li)$^5$He and $^4$He($^7$Li,$^6$He)$^5$Li stripping reactions and found a value of $a_n$ $=$ $5.5$ $\pm$ $1.0$~fm from fitting the experimental line shapes. Jarmie, Brown and Hardekopf \cite{jarmie84} and Brown, Jarmie and Hale \cite{brown87} assumed radii of $a_d$ $=$ $5.0$~fm and $a_n$ $=$ $3.0$~fm. The latter value presumably originated from Adair \cite{adair52} and Dodder and Gammel \cite{dodder52}, who adopted $a_n$ $=$ $2.9$~fm to fit the low-energy $^4$He $+$ nucleon phase shifts. In the present work, we will chose for the channel radii uniform priors between $2.5$~fm and $8.0$~fm.

The systematic uncertainty of the measured energies is treated as a (positive or negative) offset ($f_E$). The original works report total energy uncertainties only, but do not provide specific information about the relative contributions of statistical and systematic effects. We will assume that the prior, for each data set, $j$, is given by a normal density with a mean value of zero and a standard deviation equal to the average reported total energy uncertainty in that experiment (Appendix~\ref{app:data}). 

The systematic S-factor uncertainties for the data of Jarmie, Brown and Hardekopf \cite{jarmie84}, Kobzev, Salatskij and Telezhnikov \cite{kobzev66}, Arnold {\it et al.} \cite{arnold53}, and Conner, Bonner and Smith \cite{conner52} amount to $1.26$\%, $2.5$\%, $2.0$\%, and $1.8$\%, respectively (Appendix~\ref{app:data}). These correspond to factor uncertainties of $(f.u.)_1$ $=$ $1.0126$, $(f.u.)_3$ $=$ $1.025$, $(f.u.)_4$ $=$ $1.020$, and $(f.u.)_5$ $=$ $1.018$, respectively. As explained in Section~\ref{sec:likeprior}, we will use these values as shape parameters of lognormal priors for the systematic normalization factors, $f_S$, of each experiment. We already mentioned in Section~\ref{sec:data} that Brown, Jarmie and Hale \cite{brown87} did not determine absolute cross sections, but normalized their data to the results of Ref.~\cite{jarmie84}. We will include this data set in our analysis by choosing a weakly informative prior for the factor uncertainty, i.e., $(f.u.)_2$ $=$ $10$. 

Finally, the extrinsic uncertainties for both energy and S-factor are inherently unknown to the experimenter. Thus we will assume very broad truncated normal priors, with standard deviations of $10$~keV for the energy and $2$~MeVb for the S-factor. 

Our complete Bayesian model is summarized below in symbolic notation as:
\begin{align}
\label{eq:model}
    & \textrm{Parameters:} \notag \\
    & \indent \theta \equiv (E_0,E_B,\gamma^2_d,\gamma^2_n,a_d,a_n,U_e, \notag \\
    & ~~~~~~~~~~\sigma_{E,extr,j},\sigma_{S,extr,j},f_{E,j},f_{S,j}) \\     
    & \textrm{Likelihoods for energy:}\notag \\ 
    & \indent E^\prime_i  \sim N(E_i, \sigma_{E,extr,j}^2) \\
    & \indent E^{\prime\prime}_{i,j} = f_{E,j} + E^\prime_i \\
    & \indent E^{exp}_{i,j}  \sim N(E^{\prime\prime}_{i,j}, \sigma_{E,stat,i}^2) \\
    & \textrm{Likelihoods for S-factor:}\notag \\ 
    & \indent S^\prime_i  \sim N(S_i, \sigma_{S,extr,j}^2) \\
    & \indent S^{\prime\prime}_{i,j} = f_{S,j} \times S^\prime_i \\
    & \indent S^{exp}_{i,j}  \sim N(S^{\prime\prime}_{i,j}, \sigma_{S,stat,i}^2) \\
    & \textrm{Priors:}\notag \\ 
    & \indent E_0 \sim U(0.02, 0.08) 	\\
    & \indent E_B \sim N(0.0, 1.0^2) T(0,)	\\
    & \indent (\gamma_d^2, \gamma_n^2) \sim  N(0.0, (\gamma_{WL}^2)^2)T(0,)   \\
    & \indent (a_d, a_n) \sim U(2.5, 8.0) \\
    & \indent U_e \sim N(0.0, 0.001^2) T(0,)   \\
    & \indent E_i \sim U(0.001, 0.3) \\
    & \indent \sigma_{E,extr,j} \sim N(0.0, 0.01^2) T(0,) \\ 
    & \indent f_{E,j} \sim N(0.0, \xi_j^2)  \\
    & \indent \sigma_{S,extr,j} \sim N(0.0, 2.0^2) T(0,) \\ 
    & \indent f_{S,j} \sim LN(0, [ \ln(f.u.)_j ]^2)  
\end{align}
where the indices $j$ $=$ $1,...,5$ and $i$ $=$ $1,...,191$ label the data set and the data points, respectively. The symbols have the following meaning: measured energy ($E^{exp}$) and measured S-factor ($S^{exp}$); true energy ($E$); the true S-factor ($S$) is calculated from the R-matrix expressions (see Equations~(\ref{eq:eq1})$-$(\ref{eq:scatt})) using the R-matrix parameters; $N$, $U$, and $LN$ denote normal, uniform, and lognormal probability densities, respectively; $T(0,)$ indicates that the distribution is only defined for positive random variables; ``$\sim$'' stands for ``sampled from.'' The numerical values of energies, S-factors, and radii are in units of MeV, MeVb, and fm, respectively. For the standard deviation, $\xi_j$, of the prior for the systematic energy offset, $f_{E,j}$, we adopted the average value of the reported energy uncertainties for a given experiment, $j$ (see Appendix~\ref{app:data}).

\section{Results}\label{sec:results}
The MCMC sampling will provide the posteriors of all $27$ parameters. We computed three MCMC chains, where each chain had a length of 5$\times$10$^6$ steps after the burn-in samples (10$^6$ steps for each chain) were completed. The autocorrelation approached zero for a lag of $\approx$ $3000$. Therefore, the effective sample size, i.e., the number of independent Monte Carlo samples necessary to give the same precision as the actual MCMC samples, amounted to $\approx$ $5000$. This ensured that the chains reached equilibrium and Monte Carlo fluctuations were negligible compared to the statistical, systematic, and extrinsic uncertainties.

\subsection{S-factors and R-matrix parameters}
\begin{figure}[]
\includegraphics[width=1.0\columnwidth]{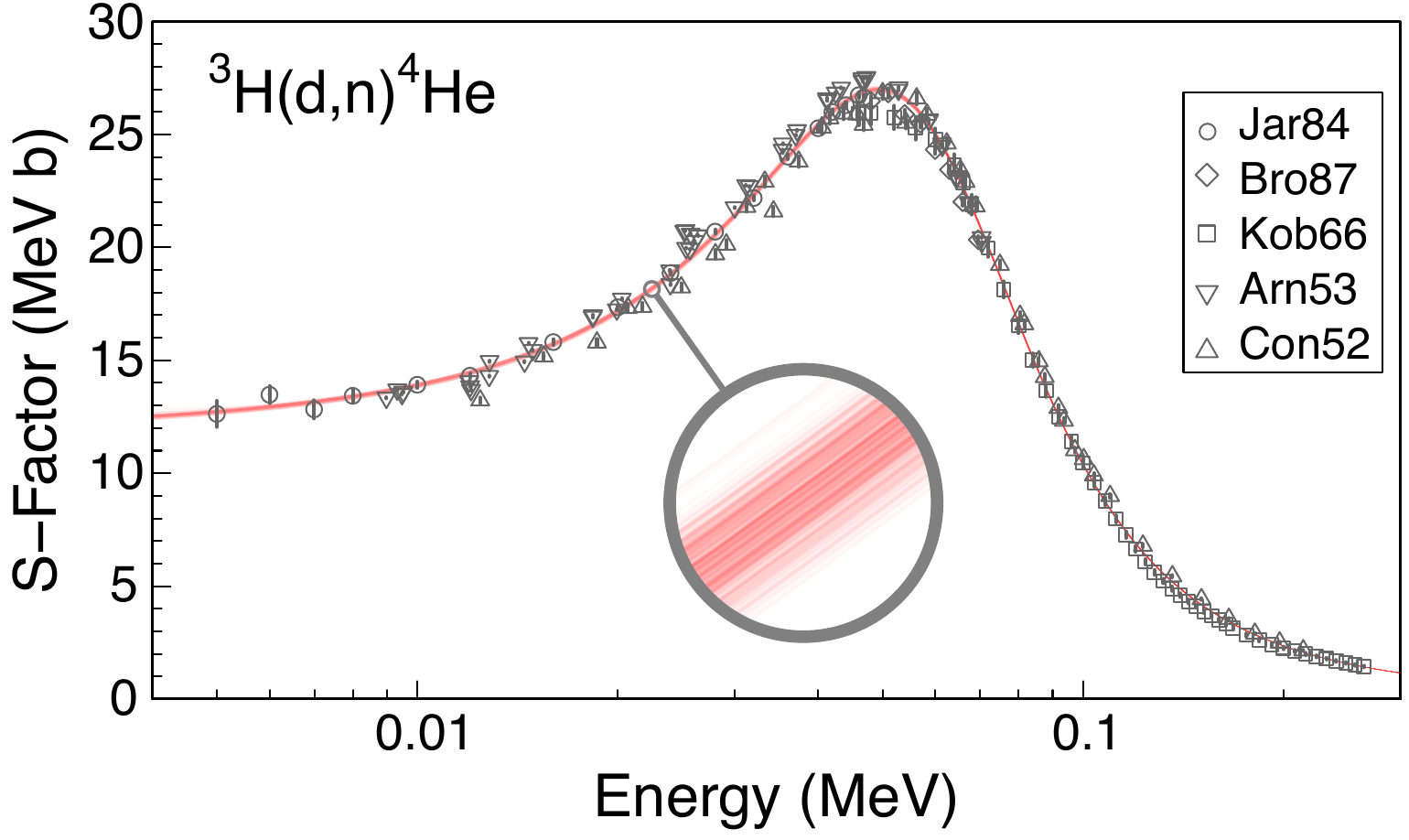}
\caption{\label{fig:BaySfac} Astrophysical S-factors obtained from the Bayesian R-matrix fit. The data are the same as in Figure~\ref{fig:data}. The red lines represent credible S-factors computed using $500$ sampled parameter sets that were chosen at random from the complete set of samples. The inset shows a magnified view of the credible S-factor samples. 
}
\end{figure}

\begin{figure}[]
\includegraphics[width=0.8\columnwidth]{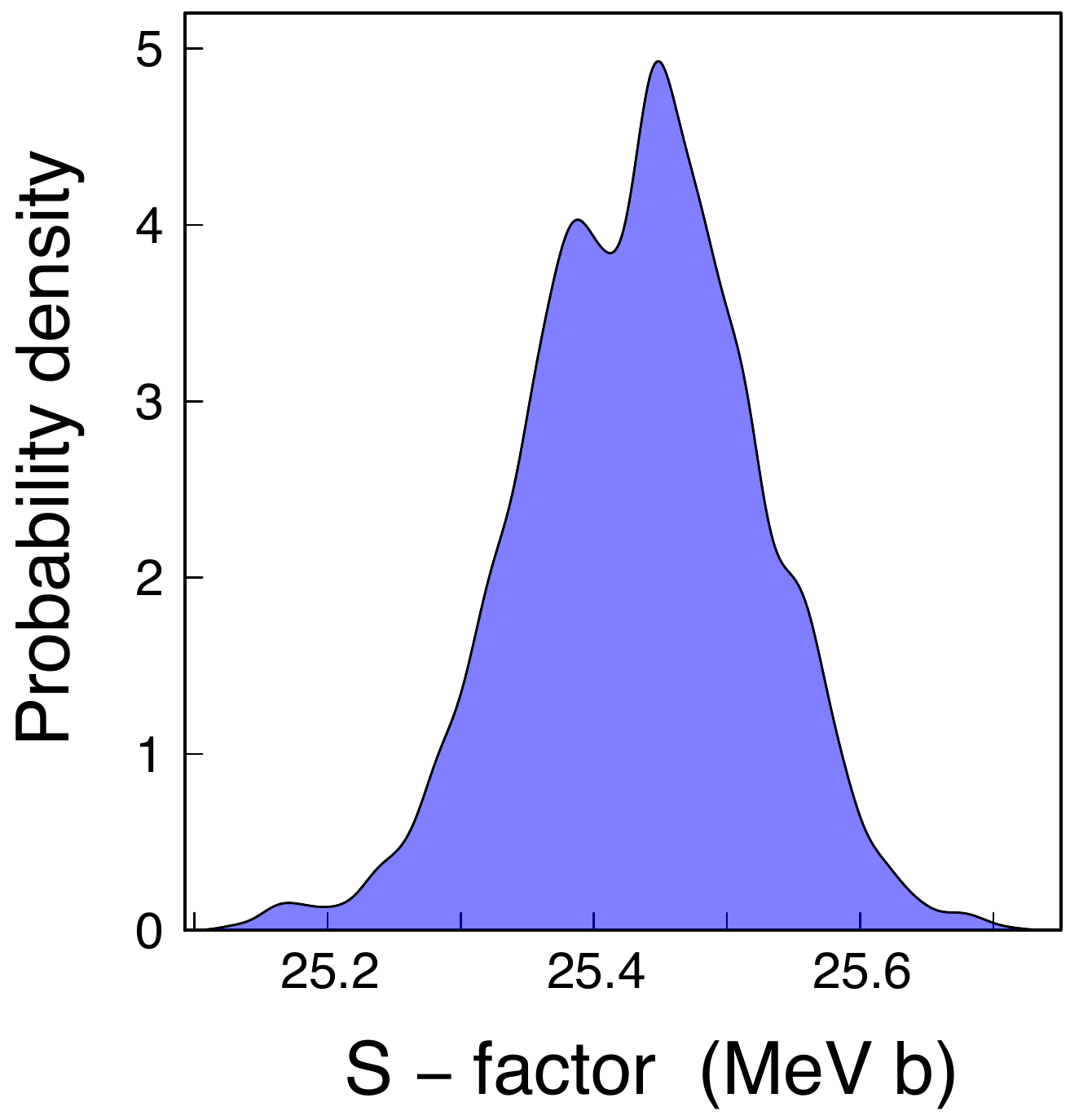}
\caption{\label{fig:sfac40} Marginalized posterior of the S-factor at a respresentative center-of-mass energy of $40$~keV. Percentiles of the distribution are listed in Table~\ref{tab:results}. 
}
\end{figure}

The results for the S-factor are displayed in Figure~\ref{fig:BaySfac}. For better visualization, the red lines represent only $500$ S-factor samples that were chosen at random from the complete set of 15$\times$10$^6$ samples. The marginalized posterior of the S-factor at a representative energy of $40$~keV, near the center of the energy range important for fuison reactors and big bang nucleosynthesis, is shown in Figure~\ref{fig:sfac40}. At this energy, we find a value of $S_{0.04}^{pres}$ $=$ 25.438$_{-0.089}^{+0.080}$~MeVb (Table~\ref{tab:results}), where the uncertainties are derived from the $16$, $50$, and $84$ percentiles. This uncertainty amounts to 0.4\%. Our result can be compared to the previous value of $S_{0.04}^{prev}$ $=$ 25.87$\pm$0.49~MeVb from Bosch and Hale \cite{bosch92}, which was obtained using different methods and data selection. The present and previous recommended values differ by 1.7\% and our uncertainty is smaller by a factor of $5.5$.
 
\begin{figure}[]
\includegraphics[width=1.0\columnwidth]{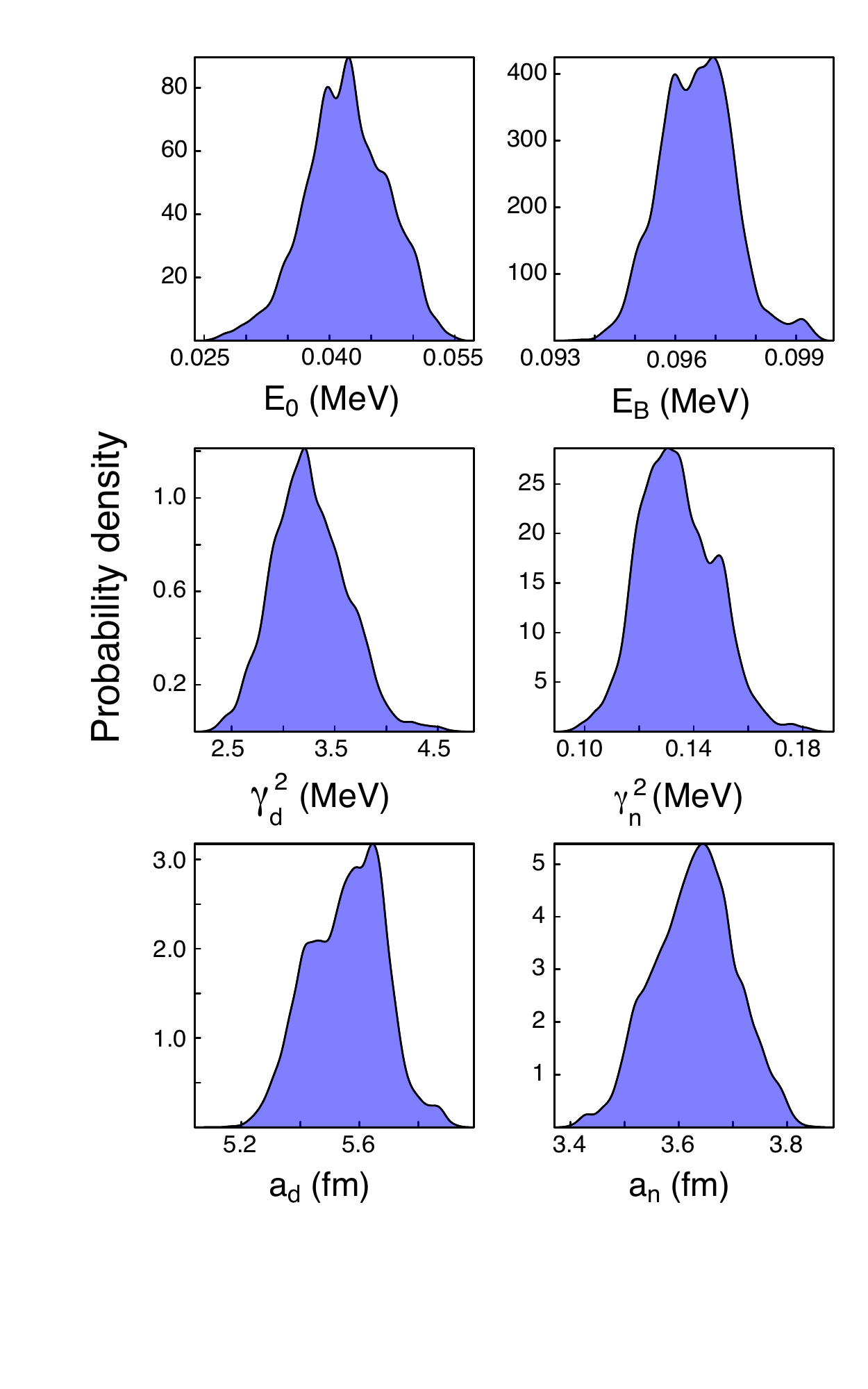}
\caption{\label{fig:rmatparam} Marginalized posterior densities of the eigenenergy ($E_0$), the energy at which the level shift is set to zero ($E_B$), the deuteron and neutron reduced widths ($\gamma_d^2$, $\gamma_n^2$), and the deuteron and neutron channel radii ($a_d$, $a_n$). Percentiles of the distributions are listed in Table~\ref{tab:results}.
}
\end{figure}

Our results for the R-matrix parameters are listed in Table~\ref{tab:results}, together with previously obtained values.  The top panels in Figure~\ref{fig:rmatparam} presents the marginalized posterior densities of the eigenenergy ($E_0$) and the energy at which the shift factor is equal to zero ($E_B$). We find values of $E_0$ $=$ 0.0420$_{-0.0047}^{+0.0051}$~MeV and $E_B$ $=$ 0.09654$_{-0.00090}^{+0.00084}$~MeV. These cannot be directly compared to the result of Barker \cite{barker97}, $0.0912$~MeV, who assumed $E_0$ $=$ $E_r$ and fixed channel radii ($a_d$ $=$ $6$~fm, $a_n$ $=$ $5$~fm) in the fit. The middle panels in Figure~\ref{fig:rmatparam} show the posteriors of the deuteron and neutron reduced widths. We obtain values of $\gamma_d^2$ $=$ 3.23$_{-0.32}^{+0.39}$~MeV, and $\gamma_n^2$ $=$ 0.133$_{-0.013}^{+0.016}$~MeV. Our deuteron reduced width agrees with Barker's result, but our neutron reduced width is larger by a factor of $1.7$. A more quantitative comparison between present and previous results is difficult, because no uncertainties are presented in Ref.~\cite{barker97}. The bottom panels in Figure~\ref{fig:rmatparam} display the posteriors of the deuteron and neutron channel radii. The present results are $a_d$ $=$ 5.56$_{-0.15}^{+0.11}$~fm and $a_n$ $=$ 3.633$_{-0.084}^{+0.072}$~fm. Our deuteron channel radius is lower than the value obtained in previous fitting \cite{argo52,hale14} (see Section~\ref{sec:tdnmodel}). Our neutron channel radius is larger than the value found previously by Refs.~\cite{adair52,dodder52}, but smaller than the results obtained in Refs.~\cite{argo52,hale14,woods88}. Again, no uncertainties are provided in the previous works. 

For completion, we also list in Table~\ref{tab:results} the values of the deuteron and neutron partial widths that are obtained from our reduced widths according to Equation~(\ref{eq:gamwidth}). We obtain best-fit values of $\Gamma_d$ $=$ 0.897$_{-0.068}^{+0.095}$~MeV and $\Gamma_n$ $=$ 0.549$_{-0.041}^{+0.055}$~MeV (Table~\ref{tab:results}). Therefore, we confirm the relation $\Gamma_d$ $\approx$ $\Gamma_n$, which explains the large cross section of the $^3$H(d,n)$^4$He reaction at low energies, as explained in Section~\ref{sec:intro}. 

\subsection{Electron screening}
Motivated by electron screening effects observed in $^3$He(d,p)$^4$He S-factor data, Langanke and Rolfs \cite{langanke89} investigated the data of Jarmie, Brown and Hardekopf \cite{jarmie84} and Brown, Jarmie and Hale \cite{brown87} of the analog $^3$H(d,n)$^4$He reaction. Based on a one-level R-matrix expression, Langanke and Rolfs \cite{langanke89} report evidence of ``electron screening effects caused by the electrons present in the target'' at the lowest center-of-mass energies ($\le$ $16$~keV). Since their R-matrix fit underpredicts the six lowest data points (see  Figure~\ref{fig:BaySfac}), they claim much better agreement if a screening potential of $41$~eV (Thomas-Fermi model) or $27$~eV (Hartree-Fock model) is included in the data fitting. 

\begin{figure}[]
\includegraphics[width=1.0\columnwidth]{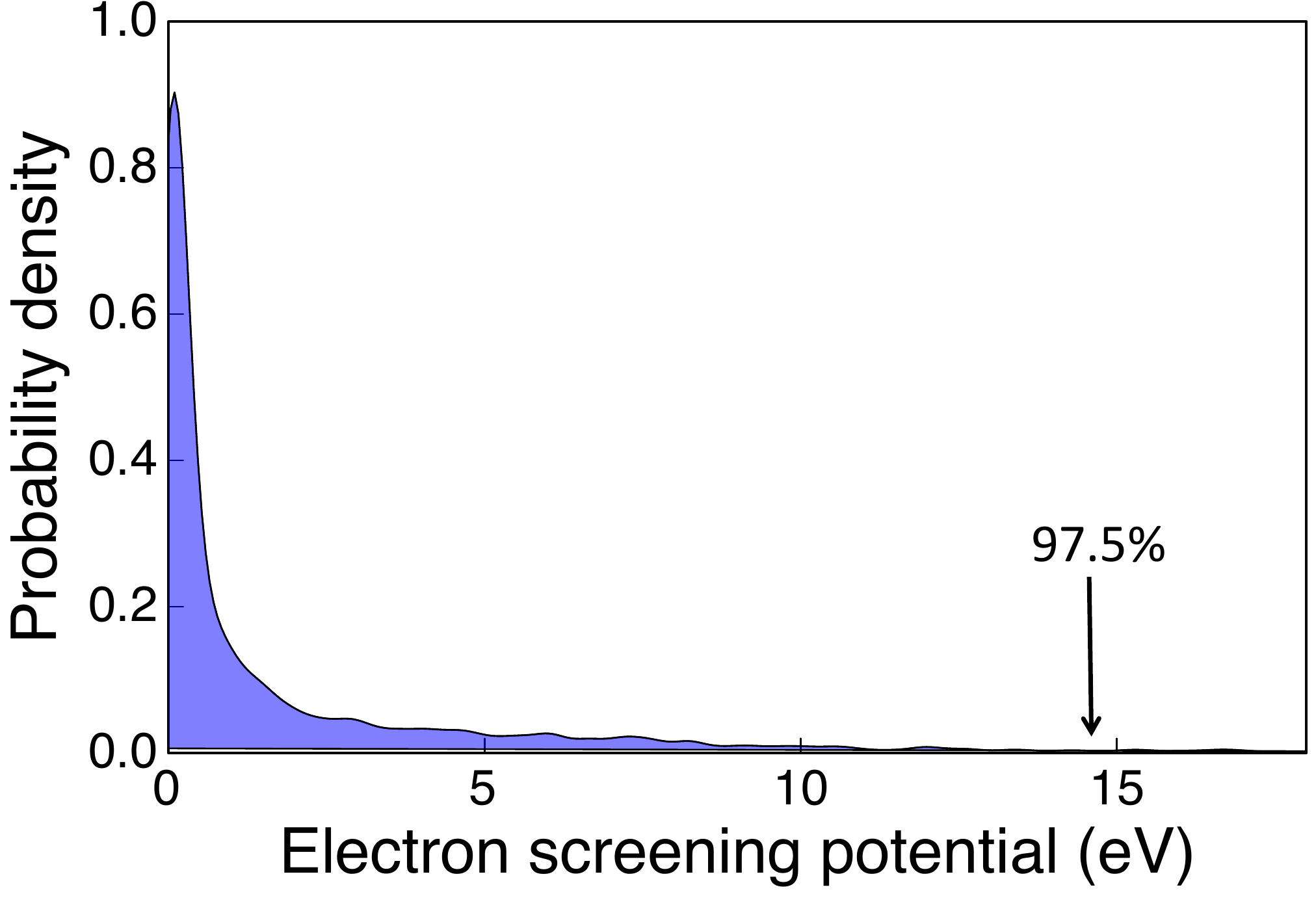}
\caption{\label{fig:eleScreen} Marginalized posterior density for the electron screening potential, $U_e$. No evidence for electron screening in the $^3$H(d,n)$^4$He reaction can be extracted from the available data, contrary to the claims of Langanke and Rolfs \cite{langanke89}, and only an upper limit, $U_e$ $\le$ $14.7$~eV, can be obtained (Table~\ref{tab:results}). 
}
\end{figure}

Figure~\ref{fig:eleScreen} shows our marginalized posterior density for the electron screening potential, $U_e$. It clearly demonstrates that there is no evidence of electron screening effects in the $^3$H(d,n)$^4$He data, and only an upper limit can be extracted from the measurements. Integration of the posterior from zero to a percentile of $97.5$\% results in an upper limit of $U_e$ $\le$ $14.7$~eV (Table~\ref{tab:results}). We suspect that the erroneous claim of electron screening effects in the $^3$H(d,n)$^4$He reaction by Langanke and Rolfs \cite{langanke89} is most likely caused by the wrong sign of the level shift in the denominator of their one-level R-matrix expression (see their Equation~4).
 
\subsection{Normalization and extrinsic scatter} 
\begin{figure}[]
\includegraphics[width=1.0\columnwidth]{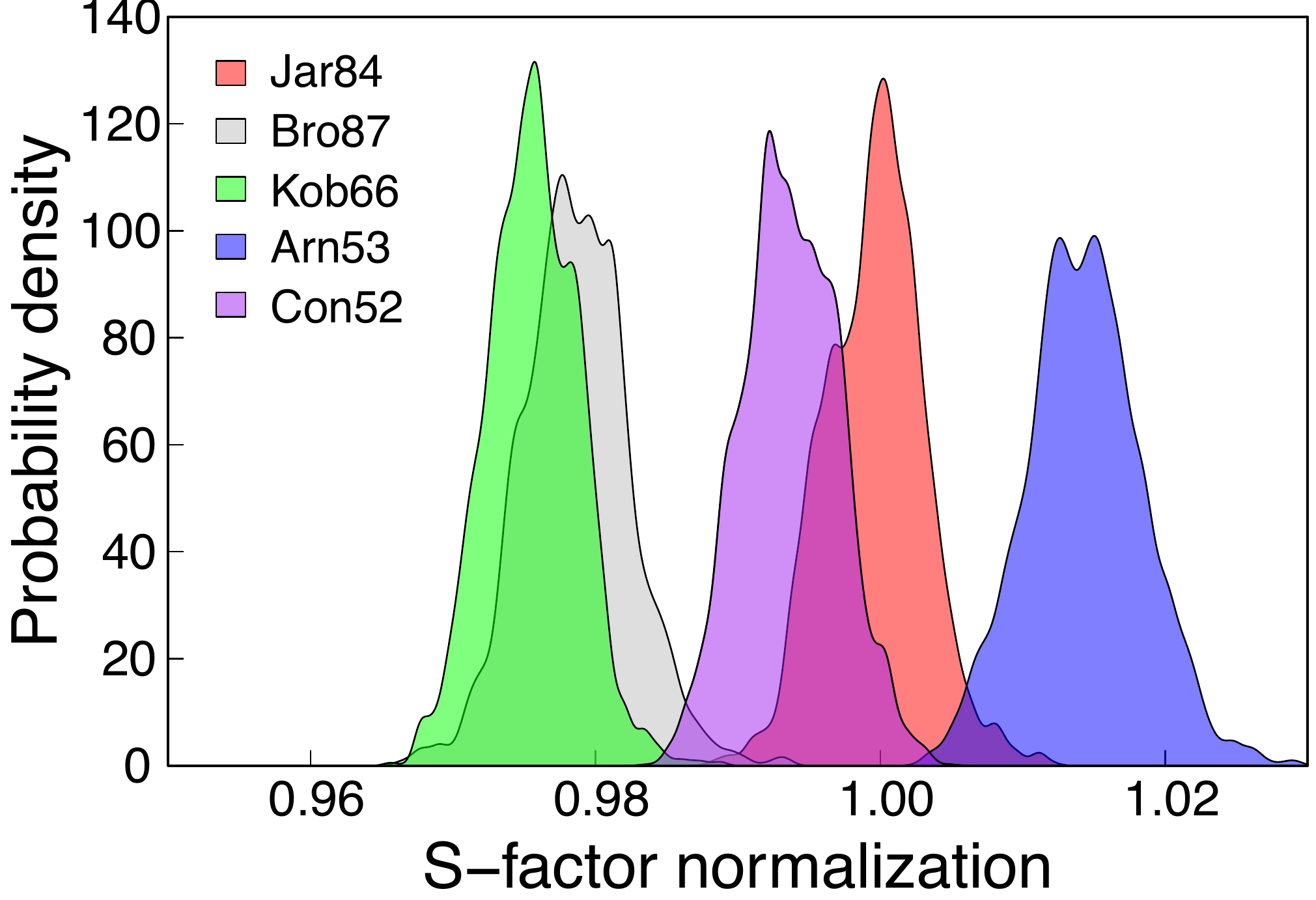}
\caption{\label{fig:normfac} Marginalized posteriors of the S-factor normalization factors, $f_S$. The labels refer to the same data sets as shown in Figure~\ref{fig:data}. Percentiles of the distributions are listed in Table~\ref{tab:results2}.
}
\end{figure}

Apart from the physical parameters discussed above, our Bayesian model also provides interesting information about systematic and extrinsic uncertainties in the data. The marginalized posteriors of the S-factor normalization factors, $f_S$, are displayed in Figure~\ref{fig:normfac}. Values for the percentiles of the distribution for each data set are listed in Table~\ref{tab:results2}. The median values of $f_S$ are equal to unity within $\approx$ $2.4$\%. They are also similar in magnitude to the factor uncertainties, $f.u.$ (Section~\ref{sec:tdnmodel}), indicating that reliable systematic S-factor uncertainties were adopted in our analysis (Appendix~\ref{app:data}). Brown and Hale \cite{brown14} find ``normalization factors'' of $1.017$ and $1.025$ for the data of Jarmie, Brown and Hardekopf \cite{jarmie84} and Brown, Jarmie and Hale \cite{brown87}, respectively, where the inverse of their value corresponds to our value of $f_S$, as explained in Section~\ref{sec:likeprior}. Our derived value, $f_S$ $=$ 0.9998$_{-0.0037}^{+0.0030}$, for the data of Ref.~\cite{jarmie84} is larger than the value of $1.017^{-1}$ $=$ $0.983$ from Brown and Hale \cite{brown14}, but our results for the data of Ref.~\cite{brown87} are in agreement.
 
Table~\ref{tab:results2} also lists the extrinsic S-factor uncertainty for each data set. The derived values can be compared with the magnitude of the statistical S-factor uncertainties, presented in Appendix~\ref{app:data}. It can be seen that for the data of Refs.~\cite{jarmie84,brown87,kobzev66} the extrinsic scatter is smaller, or of similar magnitude, compared to the reported statistical uncertainties. However, our derived extrinsic scatter for the data of Arnold {\it et al.} \cite{arnold53}, $\sigma_{S,extr}$ $=$ 0.471$_{-0.038}^{+0.043}$~MeVb, exceeds their reported statistical uncertainties by more than an order of magnitude (Table~\ref{tab:arnold}). This indicates that the latter authors underestimated their statistical uncertainties. A similar problem, but less severe, persists for the data set of Conner, Bonner and Smith \cite{conner52}.   

Regarding the energies, all of our predicted systematic shifts, $f_E$, are consistent with zero. Furthermore, for the extrinsic scatter we only find upper limits, which are smaller than the reported statistical energy uncertainties. Thus we conclude that the energies were reliably estimated in the original works.
 
Notice that even when we identify problems with certain data sets, all effects are naturally accounted for in our Bayesian model. Specifically, there is no need to arbitraily disregard data. 

\section{Thermonuclear Reaction Rates}\label{sec:rates}
In the nuclear astrophysics literature, the thermonuclear reaction rate per particle pair, $N_A \langle \sigma v \rangle$, at a given plasma temperature, $T$, is defined by \cite{iliadis15}
\begin{equation}
\begin{split}
N_A \langle \sigma v \rangle & = \left(\frac{8}{\pi m_{01}}\right)^{1/2} \frac{N_A}{(kT)^{3/2}} 
    \int_0^\infty e^{-2\pi\eta}\,S(E)\,e^{-E/kT}\,dE 
\label{eq:rate}
\end{split}
\end{equation}
where $m_{01}$ is the reduced mass of projectile and target, $N_A$ is  Avogadro's constant, and $k$ is the Boltzmann constant. In the fusion research community, the quantity $\langle \sigma v \rangle$ is called thermal reactivity and is usually presented as a function of the thermal energy, $kT$ (i.e., the maximum of the Mawell-Boltzmann {\it velocity} distribution).

We computed reaction rates and reactivities by numerical integration of Equation~(\ref{eq:rate}). The S-factor is calculated from the samples of the 27-parameter Bayesian R-matrix fit, discussed in Section~\ref{sec:results}, and thus our new values of $N_A \langle \sigma v \rangle$ and $\langle \sigma v \rangle$ fully contain the effects of varying channel radii, varying boundary condition parameters, systematic and extrinsic uncertainties. We base these results on 5,000 random MCMC S-factor samples, which ensures that Monte Carlo fluctuations are negligible compared to the reaction rate or reactivity uncertainties. Our lower integration limit was set at $1$~eV. Reaction rates are computed for $46$ different temperatures between $1$~MK and $1$~GK, and reactivities are calculated for $25$ different values of $kT$ between $0.2$~keV and $50$~keV. Recommended rates or reactivities are computed as the 50th percentile of the probability density, while the factor uncertainty, $f.u.$, is obtained from the 16th and 84th percentiles \cite{longland10}. Numerical values of reaction rates and reactivities are listed in Table~\ref{tab:rates} and \ref{tab:rates2}, respectively. 

\begin{figure}[]
\includegraphics[width=1.0\columnwidth]{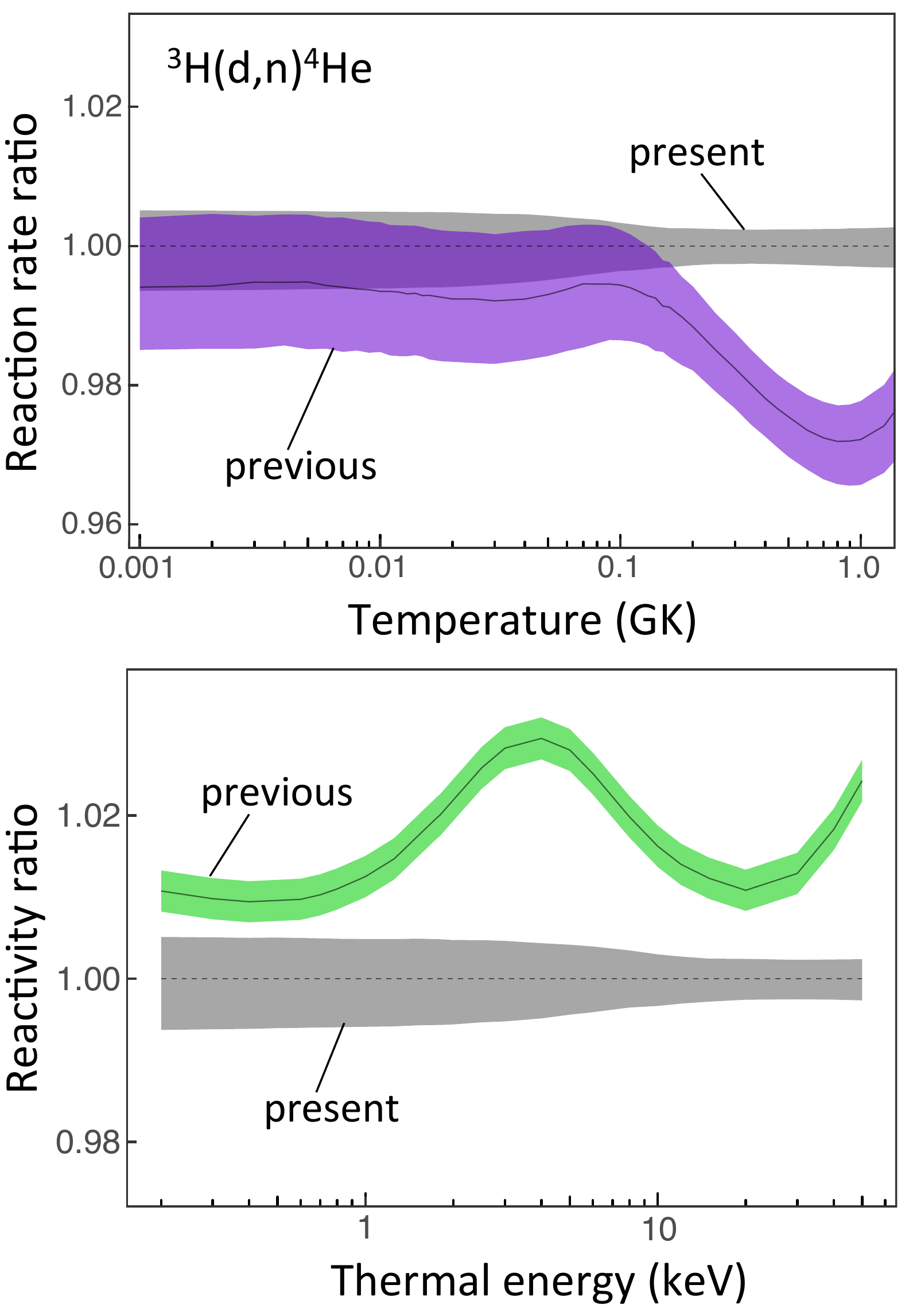}
\caption{\label{fig:ratecomp} (Top) Present $^3$H(d,n)$^4$He thermonuclear rates (gray) compared to the evaluation of Descouvemont {\it et al.} \cite{descouvemont04} (purple). (Bottom) Present $^3$H(d,n)$^4$He reactivities (gray) compared with the results of Bosch and Hale \cite{bosch92b} (green). The gray bands signify 68\% coverage probabilities. For a better comparison, all rates or reactivities are normalized to our new recommended (i.e., median) values (see Tables~\ref{tab:rates} and \ref{tab:rates2}). The solid lines shows the ratio of previous and present recommended results.
}
\end{figure}

Reaction rates are displayed in the top panel of Figure~\ref{fig:ratecomp}. Our low (16th percentile) and high (84th percentile) rates, normalized to the present median rates (50th percentile), are shown as a gray band. The rate uncertainties in the temperature region between $1$~MK and $1$~GK are between 0.2\% and 0.6\%. While a number of previous works have presented $^3$H(d,n)$^4$He thermonuclear rates, most do not present uncertainties and, therefore, a direct comparison to our results is not very meaningful. The only recently published $^3$H(d,n)$^4$He rates with uncertainties can be found in Descouvemont {\it et al.} \cite{descouvemont04}. Their ``lower'', ``adopted'', and ``upper'' rates, normalized to our median rate, are shown as the purple band in the top panel of Figure~\ref{fig:ratecomp}. Present and previous rates agree below a temperature of $0.1$~GK, although the previous rate uncertainties (0.8\% to 1.0\%), estimated using chi-square fitting, are larger compared to our results. At higher temperatures, present and previous rates start to diverge. At a temperature of 1~GK, the difference amounts to 2.9\%.

Reactivities are displayed in the bottom panel of Figure~\ref{fig:ratecomp}. Our low (16th percentile) and high (84th percentile) reactivities, normalized to the present median reactivites (50th percentile), are shown as a gray band. We compare our results with those listed in Table VIII of Bosch and Hale \cite{bosch92b}. Notice that their quoted uncertainty of 0.25\% (see Table VII in Ref.~\cite{bosch92}) has no rigorous statistical meaning but signifies the ``maximum deviation of the fit from the input data.'' The previously recommended reactivities are higher than our values at all thermal energies, with the largest deviation of 2.9\% occuring at an energy of $kT$ $=$ $4$~keV. 

\section{Summary and Conclusions}\label{sec:summary}
We presented the first Bayesian R-matrix analysis of $^3$H(d,n)$^4$He S-factors, reaction rates, and reactivities. This approach has major advantages, because it is not confined to the use of Gaussian likelihoods, and instead allows for implementing those likelihoods into the model that best apply to the problem at hand. Also, all previous R-matrix analyses kept the channel radii and boundary condition parameters constant during the fitting. In reality, these quantities are not rigidly constrained, and their variation will impact the uncertainties of the derived S-factors and fusion rates. Furthermore, uncertainties affect not only the measured S-factors, but also the experimental center-of-mass energies. Uncertainties in both independent and dependent variables can be easily implemented into a Bayesian model, whereas no simple prescription for such a procedure exists in chi-square fitting.

We evaluated the published data and adopted those experiments for which separate estimates of systematic and statistical uncertainties can be obtained: Jarmie, Brown and Hardekopf \cite{jarmie84};  Brown, Jarmie and Hale \cite{brown87}; Kobzev, Salatskij and Telezhnikov \cite{kobzev66}; Arnold {\it et al.} \cite{arnold53}; and Conner, Bonner and Smith \cite{conner52}. The difficulties and special circumstances when studying the exceptionally broad low-energy resonance in this reaction are discussed in detail. We analyzed the low-energy S-factor data using a two-channel, single-level R-matrix approximation that is implemented in a Bayesian analysis. The model has $27$ parameters, including R-matrix parameters (e.g., energies and reduced widths), systematic uncertainties, and extrinsic uncertainties. In particluar, we included in the sampling the channel radii, boundary condition parameters, and data set normalization factors. Our resulting S-factor uncertainty amounts to only 0.4\% near an energy of $40$~keV. Thermonuclear reaction rates and reactivities are found by numerically integrating the Bayesian S-factor samples. Our resulting rate or reactivity uncertainties are between 0.2\% and 0.6\%. Above $0.1$~GK, our reaction rates are larger than the values of Descouvemont {\it et al.} \cite{descouvemont04}. Our reactivities are smaller than the results of Bosch and Hale \cite{bosch92b} at all relevant thermal energies.  Finally, unlike previous claims, we find no evidence for the electron screening effect in any of the published $^3$H(d,n)$^4$He reaction data. 

The present study demonstrates the usefulness of the Bayesian approach for estimating R-matrix parameters, S-factors, reaction rates, and reactivities. The results will prove useful in future R-matrix studies that involve multiple channels and resonances.

\begin{acknowledgments}
We would like to thank Caleb Marshall for helpful comments. This work was supported in part by NASA under the Astrophysics Theory Program grant 14-ATP14-0007, and the U.S. DOE under contracts DE-FG02-97ER41041 (UNC) and DE-FG02-97ER41033 (TUNL).  
\end{acknowledgments}



\appendix
\section{Nuclear Cross Section Data for $^3$H $+$ $\lowercase{d}$ $\rightarrow$ $\lowercase{n}$ $+$ $^4$H\lowercase{e}} \label{app:data}
We discuss here the current status of the available data for the $^3$H(d,n)$^4$He reaction. Several works have measured only differential cross sections at a single angle, and assumed an isotropic angular distribution to derive the total cross section. Figure~4 in Conner, Bonner and Smith \cite{conner52} shows that the integrated cross section data points agree with the theoretical single-level dispersion curve (solid line) at deuteron bombarding energies of $\leq$ $450$~keV. Therefore, at these low energies, the cross section is determined by the $3/2^+$ (s-wave) resonance in $^3$H $+$ $d$ (see Section~\ref{sec:intro}), and the angular distribution can be assumed to be nearly isotropic; see also B\'em {\it et al.} \cite{bem97}. At higher energies, higher-lying levels in $^5$He will impact the cross section, giving rise to anisotropies in the differential cross section. In the present work, we only take data in this low-energy range into account (corresponding to bombarding triton energies of $\leq$ $680$~keV, or center-of-mass energies of $\leq$ $270$~keV), which is of primary interest for $^3$H $+$ $d$ thermonuclear fusion. As noted in Section~\ref{sec:data}, we will adopt in our analysis only those data sets for which we can separately estimate statistical and systematic uncertainties. 

\subsection{The $^2$H(t,$\alpha$)n Data of Jarmie, Brown and Hardekopf \cite{jarmie84}}
The measurement of Jarmie, Brown and Hardekopf \cite{jarmie84} was performed using a triton beam incident on a windowless deuterium gas target. This technique minimizes systematic beam energy uncertainties compared to other measurements that used a gas target contained by foils. Our adopted center-of-mass energies and astrophysical S-factors are listed in Table~\ref{tab:jarmie}. The energies (E$_{cm}$ $=$ $5$ $-$ $47$~keV) correspond to the center of the gas target and were calculated from the laboratory energies listed in column 2 of Table~V in Ref.~\cite{jarmie84}. The total (systematic plus statistical) uncertainties of the center-of-mass energies are less than $6$~eV. The S-factors are taken from column 3 of their Table~VI. Their statistical uncertainties amount to $0.5$\% $-$ $4.6$\%, depending on energy (see their Table~III). The systematic S-factor uncertainty is $1.26$\% (see their Table~IV).

\subsection{The $^3$H(d,$\alpha$)n Data of Brown, Jarmie and Hale \cite{brown87}}
The $^3$H(d,$\alpha$)n measurement of Brown, Jarmie and Hale \cite{brown87} was performed with an apparatus similar to the one described in Ref.~\cite{jarmie84}, except that a deuteron beam (E$_d$ $=$ $80$ $-$ $116$~keV) was incident on a triton gas target. However, no absolute normalization was determined in Brown, Jarmie and Hale \cite{brown87}. For the purpose of reporting their data, Ref.~\cite{brown87} determined an approximate scale by matching the cross sections in the overlapping energy region to the earlier absolute measurement of Ref.~\cite{jarmie84}. The reported astrophysical S-factors versus center-of-mass energies are listed in Table~\ref{tab:brown}. Since they represent relative results only, we implemented these data into our analysis using a weakly informative prior for the normalization factor (Section~\ref{sec:bayes}). The statistical S-factor uncertainties amount to $0.8$\%.

\subsection{The $^2$H(t,$\alpha$)n Data of Kobzev, Salatskij and Telezhnikov \cite{kobzev66}}
Kobzev, Salatskij and Telezhnikov \cite{kobzev66} measured the $^2$H(t,$\alpha$)n cross section at 90$^\circ$ in the triton bombarding energy range of E$_t$ $=$ $115$ $-$ $1650$~keV. They employed mica foils of $0.16$~mg/cm$^2$ and $0.31$~mg/cm$^2$ thickness as entrance windows of their deuterium gas target. Below a triton bombarding energy of $\approx$ $660$~keV, the differential cross section is isotropic \cite{conner52,argo52} and, therefore, we calculated the total cross section by multiplying the values listed in their table by a factor of $4\pi$. Our adopted S-factors are given in Table~\ref{tab:kobzev}. Kobzev, Salatskij and Telezhnikov \cite{kobzev66} state ``The differential cross section was measured from $115$ to $400$~keV with 2\% accuracy[,] in the range $400$ $-$ $800$~keV with 2.5\% accuracy...'' Although Kobzev, Salatskij and Telezhnikov \cite{kobzev66} do not provide separate estimates of statistical and systematic uncertanties, we will assume that the quoted values are of statistical nature. For the systematic S-factor uncertainty in their measurement, we assume a value of 2.5\%. Regarding the uncertainties in the bombarding energy, Kobzev, Salatskij and Telezhnikov \cite{kobzev66} write ``The interaction energy of tritium and deuterium nuclei was determined with 2.5\% accuracy in the range $115$ $-$ $150$~keV, with 2\% accuracy in the range $150$ $-$ $1200$~keV.....'' We adopted these uncertainties (see Table~\ref{tab:kobzev}) and assume that they refer to statistical effects. 

\subsection{The $^3$H(d,n)$^4$He Data of Arnold {\it et al.} \cite{arnold53}}
Arnold {\it et al.} \cite{arnold53} measured cross sections of the $^3$H(d,n)$^4$He reaction between $10$~keV and $120$~keV deuteron bombarding energy, using thin ($5$ $-$ $10$~$\mu$g/cm$^2$) SiO entrance foils for their tritium gas target. Their results were later published in Arnold {\it et al.}\cite{arnold54}, and Table~III in the latter paper served as the main source for their cross sections in most previous analyses; see, e.g., Ref.~\cite{angulo99}. However, Ref.~\cite{arnold54} did not report the originally measured cross sections of Arnold {\it et al.} \cite{arnold53} in their Table~III. What is listed there are energies and cross sections derived from a ``smoothed curve" based on the energy dependence of the Gamow factor. These values should not be used in fitting the data. The original data are provided in Table~VI of Ref.~\cite{arnold53}, which we adopted in our analysis. 

We disregarded the data points at the lowest deuteron bombarding energies of $7$ $-$ $11$~keV ``...because failure of the counter collimating system and excess production of condensable vapor gave good reason to expect that the experimental value of the cross sections at these energies might be low.'' Furthermore, the listed cross section values at E$_d$ $=$ $24.96$~keV, $24.91$~keV, and $24.89$~keV are certainly affected by a decimal-point error, since they are too large by one order of magnitude. Similarly, the listed cross section values at E$_d$ $=$ $49.62$~keV and $49.60$~keV are too low by one order of magnitude. Therefore, we disregarded these five data points. 

Arnold {\it et al.} \cite{arnold53} provide a detailed list of uncertainties in their Table~VIII. Statistical S-factor uncertainties amount to $0.2$\% and $0.1$\% at deuteron bombarding energies below and above $\approx$ $40$~keV, respectively. Our derived center-of-mass energies and S-factors are listed in Table~\ref{tab:arnold}. Arnold {\it et al.} \cite{arnold53} quoted systematic S-factor uncertainties (``standard error'') of $1.8$\%,  $1.5$\%, and $1.4$\% at deuteron bombarding energies of $25$~keV, $50$~keV, and $100$~keV, respectively. In the present work, we adopted a constant systematic S-factor uncertainty of $2.0$\%. The uncertainty in the center-of-mass energy is not directly stated in Ref.~\cite{arnold53}, but can be estimated based on the information provided. They write ``...at 10 keV, 100 V of change cause a 6 percent change in cross section...'' From their Table~II, considering only the S-factor uncertainties listed under ``5. Energy,'' we estimate an uncertainty of about $\pm$75~eV for the center-of-mass energy. We will adopt this value for all of their measured energies.

\subsection{The $^3$H(d,n)$^4$He Data of Conner, Bonner and Smith \cite{conner52}}
The cross section data of Conner, Bonner and Smith \cite{conner52} were obtained in two experiments, using different ion accelerators, for deuteron bombarding energies between $10$~keV and $1732$~keV. We adopted the differential cross sections measured at $90^\circ$ from their Tables I and II. We assumed an isotropic angular distribution at low energies and multiplied their differential cross section by $4\pi$ to find the total reaction cross section. Our adopted S-factors are given in Table~\ref{tab:conner}. Conner, Bonner and Smith \cite{conner52} state that the ``statistical probable error of the values from each target was about 1 percent except for the points at 10.3 and 15.4 keV.'' We disregarded these lowest energy data points because no other information is provided regarding their cross section uncertainty. For the systematic S-factor uncertainty, based on the effects of the finite solid angle, number of target atoms, and number of incident beam particles, they quote a combined uncertainty of 1.8\%. The uncertainty in the center-of-mass energy is not directly stated in Conner, Bonner and Smith \cite{conner52}, but can be estimated based on the number of significant figures shown in their Tables~I and II. We estimate an energy uncertainty of $\pm$60~eV at 12.4 keV and $\pm$600~eV at 214~keV center-of-mass energy.

\subsection{Other Data}
The following data sets were excluded from our analysis. The data of Bretscher and French \cite{bretscher49} are much smaller in magnitude compared to other data, and do not show the maximum of the resonance. The S-factor data of Jarvis and Roaf \cite{jarvis53} display an energy dependence that contradicts all other measurements; see, for example, Figure~2 in Refs.~\cite{bosch92}. The $^2$H(t,n)$^4$He measurement of Argo {\it et al.} \cite{argo52} employed relatively thick ($1.5$~mg/cm$^2$) aluminum entrance foils for their deuterium gas target. For example, tritons of 183~keV laboratory energy, after passing the entrance foil, would have lost an energy of $568$~keV in the foil, giving rise to a beam straggling of $\approx$ $31$~keV. Consequently, the uncertainties of the effective beam energy will be significant. Argo {\it et al.} \cite{argo52} stated that the beam energy {\it loss} was determined ``to within $\pm5$~keV,'' but not enough information was provided regarding the total uncertainty of the effective beam energy. Also, Argo {\it et al.} \cite{argo52} stated that their cross section data ``...have an estimated over-all accuracy of $\pm10$\%; this $\pm10$ percent arises almost entirely from the straggling and energy correction uncertainties up to energies of about $300$~keV...'' However, insufficient information is provided to disentangle the contributions of statistical and systematic effects to the total uncertainty.


%
%
\begin{table}[]
\begin{center}
\caption{Results of Bayesian fits (I) and comparison to literature. More results are listed in Table~\ref{tab:results2}.}\label{tab:results} 
\begin{ruledtabular}
\begin{tabular}{l c c}
Parameter  & Present\footnotemark[1]  & Previous  \\
\hline
$E_0$~(MeV)	   		&   0.0420$_{-0.0047}^{+0.0051}$			&	0.0912\footnotemark[2]    \\	
$E_B$~(MeV)	   		&   0.09654$_{-0.00090}^{+0.00084}$			&	0.0912\footnotemark[2]    \\	
$\gamma_d^2$~(MeV)	&   3.23$_{-0.32}^{+0.39}$ 		&      2.93\footnotemark[6]    \\
$\gamma_n^2$~(MeV)	&   0.133$_{-0.013}^{+0.016}$ 			&      0.0794\footnotemark[6]    \\
$a_d$~(fm)                       &  5.56$_{-0.15}^{+0.11}$    	&     7.0\footnotemark[3]      \\
$a_n$~(fm)                       &  3.633$_{-0.084}^{+0.072}$    		&      7.0\footnotemark[3], 5.5$\pm$1.0\footnotemark[4], 2.9\footnotemark[5]     \\
$\Gamma_d$~(MeV)        &   0.897$_{-0.068}^{+0.095}$\footnotemark[7]           &       \\
$\Gamma_n$~(MeV)        &   0.549$_{-0.041}^{+0.055}$\footnotemark[7]           &       \\
$U_e$~(eV)			& $\le$ $14.7$				  	& $41$ or $27$\footnotemark[8] \\
\hline
$S_{0.04}$ (MeVb)\footnotemark[9]    &   25.438$_{-0.089}^{+0.080}$    	&  25.87$\pm$0.49\footnotemark[10]  \\
\end{tabular}
\end{ruledtabular}
\footnotetext[1]{Uncertainties represent $16$th, $50$th, and $84$th percentiles, while upper limits correspond to $97.5$\% credibility.}
\footnotetext[2]{From Ref.~\cite{barker97}; his fit was performed with the condition $E_0$ $=$ $E_r$ and with fixed channel radii ($a_d$ $=$ $6.0$~fm, $a_n$ $=$ $5.0$~fm). No uncertainty estimates were provided.}
\footnotetext[3]{From Refs.~\cite{argo52,hale14}. No uncertainty estimates were provided, and both works assumed $a_d$ $=$ $a_n$.}
\footnotetext[4]{From Ref.~\cite{woods88}, who assumed $a_d$ $=$ $a_n$.}
\footnotetext[5]{From Refs.~\cite{adair52,dodder52}; no uncertainty estimates were provided.}
\footnotetext[6]{From Ref.~\cite{barker97}; no uncertainty estimates were provided.}
\footnotetext[7]{Calculated from the sampled reduced width values, $\gamma^2_d$ and $\gamma^2_n$, at the sampled energy values, $E_B$.}
\footnotetext[8]{From Ref.~\cite{langanke89}; the first and second value is obtained from the Thomas-Fermi model and the Hartree-Fock model, respectively.}
\footnotetext[9]{S-factor at $40$~keV.}
\footnotetext[10]{From Table~V of Ref.~\cite{bosch92}; the uncertainty of 1.9\% provided in their Table~IV has no rigorous statistical meaning, but signifies the ``maximum deviation of the approximations from the original R-matrix cross-sections.''}
\end{center}
\end{table}
%
%
\begin{table}[]
\begin{center}
\caption{Results of Bayesian fits (II). Results listed here complement those listed in Table~\ref{tab:results}.}\label{tab:results2} 
\begin{ruledtabular}
\begin{tabular}{l c}
Parameter\footnotemark[1]  & Value\footnotemark[2]    \\
\hline
$f_{E,1}$  	(eV)				& $-$0.23$_{-0.92}^{+0.79}$	   \\
$f_{E,2}$ 	(eV)				& 0.5$_{-2.3}^{+2.2}$ 	   \\
$f_{E,3}$ 	(eV)				& 81$_{-231}^{+241}$	   \\
$f_{E,4}$  	(eV)				& 3.5$_{-8.3}^{+9.3}$         \\
$f_{E,5}$	(eV)  			& 6$_{-19}^{+18}$		   \\
\hline
$\sigma_{E,extr,1}$ (eV)		& $\le$1.1			\\
$\sigma_{E,extr,2}$ (eV)		& $\le$3.0			\\
$\sigma_{E,extr,3}$ (eV)		& $\le$153		\\
$\sigma_{E,extr,4}$ (eV)		& $\le$2.9	   		\\
$\sigma_{E,extr,5}$ (eV)		& $\le$11			\\
\hline
$f_{S,1}$  					& 0.9998$_{-0.0037}^{+0.0030}$\footnotemark[3]			   \\
$f_{S,2}$ 					& 0.9786$_{-0.0036}^{+0.0035}$\footnotemark[3] 	   \\
$f_{S,3}$ 					& 0.9756$_{-0.0031}^{+0.0032}$				   \\
$f_{S,4}$  					& 1.0143$_{-0.0038}^{+0.0040}$			   \\
$f_{S,5}$  					& 0.9936$_{-0.0034}^{+0.0035}$		   \\
\hline
$\sigma_{S,extr,1}$ (MeVb)	& 0.112$_{-0.028}^{+0.048}$			   \\
$\sigma_{S,extr,2}$ (MeVb)	& 0.181$_{-0.052}^{+0.069}$			   \\
$\sigma_{S,extr,3}$ (MeVb)	& 0.0285$_{-0.0066}^{+0.0102}$			   \\
$\sigma_{S,extr,4}$ (MeVb)	& 0.471$_{-0.038}^{+0.043}$			   \\
$\sigma_{S,extr,5}$ (MeVb)	& 0.559$_{-0.053}^{+0.050}$		   \\
\end{tabular}
\end{ruledtabular}
\footnotetext[1]{The symbols $\sigma_{E,extr}$, $\sigma_{S,extr}$, $f_E$, and $f_S$ denote the extrinsic uncertainty in energy and S-factor, the systematic energy shift, and the S-factor normalization, respectively; the indices, $j$ $=$ $1...5$, label the five different data sets: (1) Jarmie, Brown and Hardekopf \cite{jarmie84}; (2) Brown, Jarmie and Hale \cite{brown87}; (3) Kobzev, Salatskij and Telezhnikov \cite{kobzev66}; (4) Arnold {\it et al.} \cite{arnold53}; (5) Conner, Bonner and Smith \cite{conner52}.}
\footnotetext[2]{Uncertainties represent $16$th, $50$th, and $84$th percentiles, while upper limits correspond to $97.5$\% credibility.}
\footnotetext[3]{Ref.~\cite{brown14} report normalization factors of $1.017$ and $1.025$ for the data of Ref.~\cite{jarmie84} and Ref.~\cite{brown87}, respectively, where their value corresponds to the inverse of our value of $f_S$ (see Section~\ref{sec:likeprior}).}
\end{center}
\end{table}
%
%
\begin{table}[h!]
\begin{center}
\caption{Recommended $^3$H(d,n)$^4$He Thermonuclear Reaction Rates, $N_A \langle \sigma v \rangle$.}\label{tab:rates} 
\begin{ruledtabular}
\begin{tabular}{lcc|lcc}
T (GK)  & Median\footnotemark[1]  & $f.u.$\footnotemark[1]   & T (GK)  & Median\footnotemark[1]  & $f.u.$\footnotemark[1]    \\
\hline
0.001 	&	1.998$\times10^{-07}$  &  1.0059	&	0.070 	 &	1.527$\times10^{+07}$ &  1.0041	\\
0.002 	&	1.445$\times10^{-03}$  &  1.0058	&	0.080 	 &	2.348$\times10^{+07}$ &  1.0039	\\
0.003 	&	1.046$\times10^{-01}$  &  1.0058 	&	0.090 	 &	3.356$\times10^{+07}$ &  1.0037	\\	
0.004 	 &	1.539$\times10^{+00}$ &  1.0057 	&	0.100 	 &	4.536$\times10^{+07}$ &  1.0035	\\
0.005 	 &	1.034$\times10^{+01}$ &  1.0057	&	0.110 	 &	5.866$\times10^{+07}$ &  1.0033	\\
0.006 	 &	4.405$\times10^{+01}$ &  1.0056	&	0.120 	 &	7.320$\times10^{+07}$ &  1.0032	\\	
0.007 	 &	1.397$\times10^{+02}$ &  1.0056	&	0.130 	 &	8.872$\times10^{+07}$ &  1.0031	\\
0.008 	 &	3.614$\times10^{+02}$ &  1.0056	&	0.140 	 & 	 1.050$\times10^{+08}$  &  1.0030	\\
0.009 	 &	8.060$\times10^{+02}$ &  1.0056	&	0.150 	 & 	 1.217$\times10^{+08}$  &  1.0029	\\
0.010 	 &	1.606$\times10^{+03}$ &  1.0055	&	0.160 	 &	 1.388$\times10^{+08}$  &  1.0028	\\
0.011 	 &	2.934$\times10^{+03}$ &  1.0055	&	0.180 	 &	 1.732$\times10^{+08}$  &  1.0027	\\
0.012 	 &	4.998$\times10^{+03}$ &  1.0055	&	0.200 	 &	 2.069$\times10^{+08}$  &  1.0026	\\
0.013 	 &	8.044$\times10^{+03}$ &  1.0054	&	0.250 	 &	 2.843$\times10^{+08}$  &  1.0025	\\
0.014 	 &	1.235$\times10^{+04}$ &  1.0054	&	0.300 	 &	 3.483$\times10^{+08}$  &  1.0024	\\
0.015 	 &	1.824$\times10^{+04}$ &  1.0054	&	0.350 	 &	 3.988$\times10^{+08}$  &  1.0024	\\
0.016 	 &	2.604$\times10^{+04}$ &  1.0054	&	0.400 	 &	 4.375$\times10^{+08}$  &  1.0024	\\
0.018 	 &	4.891$\times10^{+04}$ &  1.0053	&	0.450 	 &	 4.663$\times10^{+08}$  &  1.0025	\\	
0.020 	 &	8.416$\times10^{+04}$ &  1.0053	&	0.500 	 &	 4.873$\times10^{+08}$  &  1.0025	\\	
0.025 	 &	2.499$\times10^{+05}$ &  1.0052	&	0.600 	 &	 5.119$\times10^{+08}$  &  1.0026	\\	
0.030 	 &	5.743$\times10^{+05}$ &  1.0050	&	0.700 	 &	 5.210$\times10^{+08}$  &  1.0026	\\	
0.040 	 &	1.942$\times10^{+06}$ &  1.0048	&	0.800 	 &	 5.206$\times10^{+08}$  &  1.0027	\\	
0.050 	 &	4.638$\times10^{+06}$ &  1.0046	&	0.900 	 &	 5.145$\times10^{+08}$  &  1.0028	\\	
0.060 	 &	9.013$\times10^{+06}$ &  1.0043	&	1.000 	 &	 5.050$\times10^{+08}$  &  1.0028	\\			
\end{tabular}
\end{ruledtabular}
\footnotetext[1]{Reaction rates in units of cm$^3$ mol$^{-1}$ s$^{-1}$, corresponding to the 50th percentile of the rate probability density function. The rate factor uncertainty, $f.u.$, is obtained from the 16th and 84th percentiles (see the text).}
\end{center}
\end{table}
%
%
\begin{table}[h!]
\begin{center}
\caption{Recommended $^3$H(d,n)$^4$He Reactivities, $\langle \sigma v \rangle$.}\label{tab:rates2} 
\begin{ruledtabular}
\begin{tabular}{lcc|lcc}
kT (keV)  & Median\footnotemark[1]  & $f.u.$\footnotemark[1]   & kT (keV)  & Median\footnotemark[1]  & $f.u.$\footnotemark[1]    \\
\hline
0.2  &   1.241$\times10^{-26}$  &    1.0058  &   3.0  &    1.816$\times10^{-18}$  &    1.0049  \\
0.3  &   7.221$\times10^{-25}$  &    1.0057  &  4.0  &     5.803$\times10^{-18}$  &    1.0046  \\
0.4  &   9.257$\times10^{-24}$  &    1.0057  &  5.0  &     1.329$\times10^{-17}$  &    1.0043  \\
0.5  &   5.643$\times10^{-23}$  &    1.0056  &  6.0  &     2.491$\times10^{-17}$  &    1.0040  \\
0.6  &   2.231$\times10^{-22}$  &    1.0056  &  8.0  &     6.101$\times10^{-17}$  &    1.0036  \\
0.7  &   6.671$\times10^{-22}$  &    1.0056  &  10.0  &   1.118$\times10^{-16}$  &    1.0032  \\  
0.8  &   1.644$\times10^{-21}$  &    1.0055  &  12.0  &   1.723$\times10^{-16}$  &    1.0029  \\
1.0  &   6.772$\times10^{-21}$  &    1.0054  & 15.0  &    2.707$\times10^{-16}$  &    1.0027  \\ 
1.3  &   3.126$\times10^{-20}$  &    1.0054  &  20.0  &   4.284$\times10^{-16}$  &    1.0025  \\
1.5  &   6.805$\times10^{-20}$  &    1.0053  &  30.0  &   6.596$\times10^{-16}$  &    1.0024  \\
1.8  &   1.738$\times10^{-19}$  &    1.0052  &  40.0  &   7.854$\times10^{-16}$  &    1.0025  \\
2.0  &   2.913$\times10^{-19}$  &    1.0052  &  50.0  &   8.444$\times10^{-16}$  &    1.0026  \\
2.5  &   8.212$\times10^{-19}$  &    1.0050  &           &                           &        \\
\end{tabular}
\end{ruledtabular}
\footnotetext[1]{Reactivities in units of cm$^3$ s$^{-1}$, corresponding to the 50th percentile of the rate probability density function. The rate factor uncertainty, $f.u.$, is obtained from the 16th and 84th percentiles (see the text).}
\end{center}
\end{table}
%
%
\begin{table}[h!]
\begin{center}
\caption{The $^2$H(t,$\alpha$)n Data of Jarmie, Brown and Hardekopf \cite{jarmie84}.}\label{tab:jarmie} 
\begin{ruledtabular}
\begin{tabular}{rc|cc}
$E_{cm}$\footnotemark[1]  & $S \pm \Delta S_{\mathrm{stat}}\footnotemark[2]$  & $E_{cm}$\footnotemark[1]   & $S \pm \Delta S_{\mathrm{stat}}\footnotemark[2]$  \\
(keV)  &  (MeVb) &  (keV)  &  (MeVb) \\
\hline
      4.992   &   12.63$\pm$0.58  	&  27.996   &   20.70$\pm$0.09  \\
      5.990   &   13.48$\pm$0.39  	&  31.998   &   22.19$\pm$0.11  \\
      6.990   &   12.83$\pm$0.40  	&  36.001   &   24.02$\pm$0.11  \\
      7.990   &   13.43$\pm$0.27  	&  40.004   &   25.28$\pm$0.14  \\
      9.989   &   13.92$\pm$0.14 		&  42.005   &   26.00$\pm$0.12  \\
     11.989   &   14.32$\pm$0.10  	&  44.007   &   26.30$\pm$0.14  \\
     15.990   &   15.81$\pm$0.13 	&  46.009   &   26.74$\pm$0.13  \\
     19.992   &   17.35$\pm$0.09  	&  46.809   &   26.64$\pm$0.14  \\
     23.994   &   18.87$\pm$0.08  	&     &     \\
\end{tabular}
\end{ruledtabular}
\footnotetext[1]{Total uncertainty varies from $\pm$2.4~eV at E$_{cm}$ $=$ $5$~keV to $\pm$6.4~eV at E$_{cm}$ $=$ $47$~keV.}
\footnotetext[2]{Systematic uncertainty: 1.26\%.}
\end{center}
\end{table}
%
%
\begin{table}[h!]
\begin{center}
\caption{The $^3$H(d,$\alpha$)n Data of Brown, Jarmie and Hale \cite{brown87}.}\label{tab:brown} 
\begin{ruledtabular}
\begin{tabular}{rc|cc}
$E_{cm}$\footnotemark[1]  & $S_{\mathrm{rel}} \pm \Delta S_{\mathrm{stat}}\footnotemark[2]$  & $E_{cm}$\footnotemark[1]   & $S_{\mathrm{rel}} \pm \Delta S_{\mathrm{stat}}\footnotemark[2]$  \\
(keV)  &  (MeVb) &  (keV)  &  (MeVb) \\
\hline
    47.948   &   26.48$\pm$0.21  &      59.941   &   24.33$\pm$0.19  \\
    50.947   &   26.84$\pm$0.21  &      62.941   &   23.44$\pm$0.19  \\
    53.942   &   25.89$\pm$0.21  &      65.941   &   22.02$\pm$0.18  \\
    56.942   &   25.50$\pm$0.20  &      69.541   &   20.34$\pm$0.16  \\
\end{tabular}
\end{ruledtabular}
\footnotetext[1]{Total uncertainty of center-of-mass energy is $\pm$9~eV.}
\footnotetext[2]{The values reported in Ref.~\cite{brown87} were normalized relative to the data of Ref.~\cite{jarmie84}, listed in Table~\ref{tab:jarmie}.}
\end{center}
\end{table}
%
%
\begin{table}[h!]
\begin{center}
\caption{The $^2$H(t,$\alpha$)n Data of Kobzev, Salatskij and Telezhnikov \cite{kobzev66}.}\label{tab:kobzev} 
\begin{ruledtabular}
\begin{tabular}{cc|cc}
$E_{cm} \pm \Delta E_{cm}$\footnotemark[1]  & $S \pm \Delta S_{\mathrm{stat}}\footnotemark[2]$  & $E_{cm} \pm \Delta E_{cm}$\footnotemark[1]  & $S \pm \Delta S_{\mathrm{stat}}\footnotemark[2]$  \\
(keV)  &  (MeVb)  & (keV)  &  (MeVb) \\
\hline
    46.0$\pm$1.2   &    25.93$\pm$0.52   &      132.0$\pm$2.6   &     5.23$\pm$0.10   \\
    48.0$\pm$1.2   &    25.96$\pm$0.52   &      136.0$\pm$2.7   &     4.89$\pm$0.10  \\
    52.0$\pm$1.3   &   25.76$\pm$0.52    &      140.0$\pm$2.8   &     4.60$\pm$0.09  \\
    56.0$\pm$1.4   &    25.28$\pm$0.51   &      144.0$\pm$2.9   &     4.32$\pm$0.09   \\
    60.0$\pm$1.5   &   24.77$\pm$0.50    &      148.0$\pm$3.0   &     4.11$\pm$0.08   \\
    64.0$\pm$1.3   &    23.66$\pm$0.47   &      152.0$\pm$3.0   &     3.88$\pm$0.08   \\
    66.0$\pm$1.3   &    22.85$\pm$0.46   &      156.0$\pm$3.1   &     3.69$\pm$0.07   \\
    68.0$\pm$1.4   &    21.89$\pm$0.44   &      160.0$\pm$3.2   &     3.50$\pm$0.07   \\
    72.0$\pm$1.4   &    19.98$\pm$0.40   &      164.0$\pm$3.3   &     3.32$\pm$0.08   \\
    76.0$\pm$1.5   &    18.14$\pm$0.36   &      168.0$\pm$3.4   &     3.15$\pm$0.08   \\
    80.0$\pm$1.6   &    16.53$\pm$0.33   &      176.0$\pm$3.5   &     2.84$\pm$0.07   \\
    84.0$\pm$1.7   &    15.01$\pm$0.30   &      184.0$\pm$3.7   &     2.62$\pm$0.07   \\
    88.0$\pm$1.8   &    13.65$\pm$0.27   &      192.0$\pm$3.8   &     2.42$\pm$0.06   \\
    92.0$\pm$1.8   &    12.50$\pm$0.25   &      200.0$\pm$4.0   &     2.26$\pm$0.06   \\
    96.0$\pm$1.9   &    11.41$\pm$0.23   &      208.0$\pm$4.2   &     2.13$\pm$0.05   \\
   100.0$\pm$2.0   &    10.45$\pm$0.21  &      216.0$\pm$4.3   &     2.00$\pm$0.05   \\
   104.0$\pm$2.1   &     9.59$\pm$0.19   &      224.0$\pm$4.5   &     1.89$\pm$0.05   \\
   108.0$\pm$2.2   &     8.76$\pm$0.18   &      232.0$\pm$4.6   &     1.79$\pm$0.04   \\
   112.0$\pm$2.2   &     7.98$\pm$0.16   &      240.0$\pm$4.8   &     1.69$\pm$0.04   \\
   116.0$\pm$2.3   &     7.28$\pm$0.15   &      248.2$\pm$5.0   &     1.60$\pm$0.04   \\
   120.0$\pm$2.4   &     6.65$\pm$0.13   &      256.2$\pm$5.1   &     1.51$\pm$0.04    \\
   124.0$\pm$2.5   &     6.08$\pm$0.12   &      264.3$\pm$5.3   &     1.44$\pm$0.04   \\
   128.0$\pm$2.6   &     5.61$\pm$0.11   &         &    \\
\end{tabular}
\end{ruledtabular}
\footnotetext[1]{Triton laboratory energies have a 2.5\% accuracy in the range $115$ $-$ $150$~keV, and a 2\% accuracy in the range $150$ $-$ $1200$~keV (see text).}
\footnotetext[2]{Assumed systematic uncertainty: 2.5\% (see text).}
\end{center}
\end{table}
%
%
\begin{table}[h!]
\begin{center}
\caption{The $^3$H(d,n)$^4$He Data of Arnold {\it et al.} \cite{arnold53}.}\label{tab:arnold} 
\begin{ruledtabular}
\begin{tabular}{rc|cc}
$E_{cm}$\footnotemark[1]  & $S \pm \Delta S_{\mathrm{stat}}\footnotemark[2]$  & $E_{cm}$\footnotemark[1]   & $S \pm \Delta S_{\mathrm{stat}}\footnotemark[2]$  \\
(keV)  &  (MeVb) &  (keV)  &  (MeVb) \\
\hline
8.98	   &	13.340$\pm$0.026   &	31.52   &	22.695$\pm$0.023  \\				
9.32	   &	13.703$\pm$0.027   &	35.36   &	24.314$\pm$0.024  \\			
9.47	   &	13.508$\pm$0.027   &	35.38   &	24.589$\pm$0.024  \\			
9.52	   &	13.600$\pm$0.027   &	37.00   &	24.967$\pm$0.025  \\				
11.95   &	14.068$\pm$0.028   &	37.16   &	25.184$\pm$0.025  \\				
11.99   &	13.849$\pm$0.028   &	41.23   &	26.600$\pm$0.027  \\				
12.03   &	13.680$\pm$0.027   &	41.25   &	26.514$\pm$0.026  \\				
12.81   &	14.302$\pm$0.029   &	43.29   &	27.067$\pm$0.027  \\				
12.83   &	14.957$\pm$0.030   &	42.49   &	26.847$\pm$0.027  \\				
14.48   &	14.939$\pm$0.030   &	46.61   &	27.466$\pm$0.027  \\				
14.68   &	15.753$\pm$0.031   &	46.64   &	27.365$\pm$0.027  \\				
14.89   &	15.448$\pm$0.030   &	46.65   &	27.489$\pm$0.027  \\				
18.33   &	16.921$\pm$0.034   &	47.22   &	27.505$\pm$0.027  \\				
18.35   &	16.989$\pm$0.032   &	47.25   &	27.542$\pm$0.027  \\				
19.92   &	17.249$\pm$0.034   &	52.80   &	26.975$\pm$0.027  \\				
20.27   &	17.721$\pm$0.035   &	52.83   &	27.085$\pm$0.027  \\				
23.95   &	18.969$\pm$0.038   &	58.66   &	25.621$\pm$0.025  \\				
23.97   &	18.366$\pm$0.036   &	58.68   &	25.669$\pm$0.026  \\				
25.17   &	20.718$\pm$0.021   &	61.39   &	24.593$\pm$0.024  \\				
25.26   &	20.755$\pm$0.021   &	61.43   &	24.492$\pm$0.024  \\				
25.32   &	19.969$\pm$0.020   &	64.51   &	23.071$\pm$0.023  \\				
25.66   &	19.920$\pm$0.020   &	64.54   &	23.157$\pm$0.023  \\				
25.72   &	20.596$\pm$0.020   &	67.37   &	22.002$\pm$0.022  \\				
26.09   &	20.277$\pm$0.020   &	67.39   &	21.951$\pm$0.022  \\				
26.38   &	20.525$\pm$0.020   &	70.39   &	20.445$\pm$0.020  \\				
29.95   &	21.766$\pm$0.022   &	70.44   &	20.227$\pm$0.020  \\				
31.16   &	22.749$\pm$0.023   &		    &				       \\			
\end{tabular}
\end{ruledtabular}
\footnotetext[1]{Total uncertainty of center-of-mass energy is about $\pm$75~eV (see text).}
\footnotetext[2]{Adopted systematic uncertainty: 2.0\% (see text).}
\end{center}
\end{table}
%
%
\begin{table}[h!]
\begin{center}
\caption{The $^3$H(d,n)$^4$He Data of Conner, Bonner and Smith \cite{conner52}.}\label{tab:conner} 
\begin{ruledtabular}
\begin{tabular}{rc|cc}
$E_{cm}$\footnotemark[1]  & $S \pm \Delta S_{\mathrm{stat}}$\footnotemark[2]  & $E_{cm}$\footnotemark[1]   & $S \pm \Delta S_{\mathrm{stat}}$\footnotemark[2]  \\
(keV)  &  (MeVb) &  (keV)  &  (MeVb) \\
\hline
    12.42   &   13.23$\pm$0.13   &       65.40   &   23.43$\pm$0.23   \\  
    15.48   &   15.17$\pm$0.15   &       66.60   &   22.90$\pm$0.23    \\    
    18.60   &   15.79$\pm$0.16   &       69.00   &   21.82$\pm$0.22    \\ 
    20.70   &   17.33$\pm$0.17   &      75.00   &   19.23$\pm$0.20    \\   
    21.78   &   17.38$\pm$0.17   &      80.40   &   16.97$\pm$0.17    \\   
    24.90   &   18.23$\pm$0.18   &      81.60   &   16.60$\pm$0.17    \\     
    28.02   &   19.70$\pm$0.20   &     85.80   &   14.96$\pm$0.15    \\    
    29.10   &   20.13$\pm$0.20   &     87.60   &   14.27$\pm$0.14    \\    
    31.20   &   21.80$\pm$0.22   &     91.80   &   12.90$\pm$0.13    \\   
    33.24   &   22.91$\pm$0.23   &     93.60   &   12.33$\pm$0.12    \\   
    34.26   &   21.59$\pm$0.21   &     97.20   &   11.02$\pm$0.11   \\    
    37.38   &   23.80$\pm$0.24   &     100.2  &  10.63$\pm$0.11    \\   
    40.50   &   25.31$\pm$0.25   &     103.8  &   9.91$\pm$0.10    \\   
    41.58   &   25.72$\pm$0.26   &      109.8  &    8.99$\pm$0.09    \\  
    43.68   &   25.93$\pm$0.26   &      123.0  &    6.79$\pm$0.07    \\   
    45.72   &   25.90$\pm$0.26   &       136.2  &    5.44$\pm$0.05    \\ 
    46.80   &   25.44$\pm$0.25   &     150.6  &    4.43$\pm$0.04    \\   
    49.98   &   26.83$\pm$0.27   &     165.6  &    3.55$\pm$0.04    \\ 
    54.18   &   25.53$\pm$0.26   &      181.2  &    2.89$\pm$0.03    \\ 
    56.22   &   26.60$\pm$0.27   &       197.4  &    2.51$\pm$0.03    \\
    58.26   &   25.89$\pm$0.26   &      214.2  &    2.16$\pm$0.02    \\
    62.40   &   24.61$\pm$0.25   &    \\
\end{tabular}
\end{ruledtabular}
\footnotetext[1]{We assumed that the uncertainty varies from $\pm$60~eV at 12.4 keV to $\pm$600~eV at 214~keV center-of-mass energy (see text).}
\footnotetext[2]{Adopted systematic uncertainty: 1.8\% (see text).}
\end{center}
\end{table}

\bibliography{paper.bib}
\end{document}